\documentclass[fleqn,usenatbib]{mnras}

\usepackage{newtxtext,newtxmath}

\usepackage[T1]{fontenc}

\DeclareRobustCommand{\VAN}[3]{#2}
\let\VANthebibliography\thebibliography
\def\thebibliography{\DeclareRobustCommand{\VAN}[3]{##3}\VANthebibliography}

\usepackage{graphicx}	
\usepackage{amsmath}	

\newcommand{\hii}{H\textsc{ii}}
\def\ks{km\,s$^{-1}$}

\def\m{$^\prime$}
\def\s{$^{\prime\prime}$}

\def\cm3{cm$^{-3}$}

\def\2{$^{12}$CO}
\def\3{$^{13}$CO}
\def\8{C$^{18}$O}

\def\msol{M$_\odot$}

\def\cm2{cm$^{-2}$}

\title[Unveiling the nature of 4FGL~J1846.9-0227]{A comprehensive analysis towards the Fermi-LAT source 4FGL J1846.9$-$0227: jets of a proto-planetary nebula producing $\gamma$-rays?}

\author[M. E. Ortega et al.]{
M. E. Ortega,$^{1}$\thanks{E-mail: mortega@iafe.uba.ar}
A. Petriella,$^{1}$
S. Paron$^{1}$
\\
$^{1}$CONICET-Universidad de Buenos Aires. Instituto de Astronom\'{\i}a y F\'{\i}sica del Espacio,
            Ciudad Universitaria, \\ (C1428EGA) Ciudad Autónoma de Buenos Aires, Argentina.\\
}

\date{Accepted XXX. Received YYY; in original form ZZZ}

\pubyear{2024}

\begin{document}
\label{firstpage}
\pagerange{\pageref{firstpage}--\pageref{lastpage}}
\maketitle

\begin{abstract}
Most of the $\gamma$-ray sources in the Fermi-LAT 14-year Source Catalogue are associated with pulsars and blazars. However, unveiling the nature of the still unassociated $\gamma$-ray sources is important for the understanding of high energy emission mechanisms in astrophysical objects. 
This work presents a comprehensive study towards the region covered by the Fermi source 4FGL J1846.9$-$0227, previously suggested to be a blazar and a massive protostar.  Using multiwavelength observations, we analysed several astrophysical objects in the region as possible counterparts of the Fermi-LAT source. Having discarded most of them after a detailed and comprehensive analysis, we suggest that the most likely candidate to be such a counterpart is IRAS 18443$-$0231, a likely proto-planetary nebula.
We found that the radio continuum emission at 3 GHz of IRAS 18443$-$0231 shows a compact source related to faint emission with jet-like morphology. Additionally, we identified an associated red-shifted CO molecular outflow. Using data from several catalogues, we obtained  radio spectral index values ranging from $-$0.57 to $-$0.39 for IRAS 18443$-$0231, indicating syncrothron emission due to particles accelerated by the jets. We point out that these jets could explain the $\gamma$-ray emission through mechanisms such as proton-proton collisions and relativistic Bremsstrahlung.
IRAS 18443$-$0231, lying almost at the centre of the Fermi confidence ellipse and related to the hard X-ray source 4XMM J184700.4$-$022752, would be the first association between a proto-planetary nebula and $\gamma$-ray emission.

\end{abstract}




\section{Introduction}
\label{intro}

Fermi-LAT unassociated sources represent some of the most enigmatic $\gamma$-ray sources in the sky. 
In the latest data release of the Fermi Gamma-ray Space Telescope (the 4th Fermi LAT 14 yr Catalogue, or 4FGL-DR4), containing 7194 $\gamma$-ray sources \citep{ballet2023}, more than 50\% of the suspected Galactic sources (i.e., $\lvert b \rvert \leq 5^{\circ}$) are yet to be identified \citep{rangelov2024}.

The nature of the $\gamma$-ray sources may be varied, they can be pulsar-wind nebulae (PWNe) and supernova remnants (SNRs) (e.g. \citealt{acero13} and \citealt{cao23}), active galactic nuclei (AGNs; e.g. \citealt{made16}),  massive young stellar objects (MYSOs; e.g. \citealt{deona23}), X-ray binaries (XRBs; e.g. \citealt{kret19} and \citealt{bah2023}), and classic- and symbiotic-novae \citep{fran2018}.  According to the 4FGL catalogues, many of the GeV-emitting PWNe and SNRs are found to be extended LAT sources, many AGNs are found to be variable sources, classical and symbiotic novae are only found to be GeV emitters during outbursts, and MYSOs were argued to be GeV emitters based mainly on positional coincidences.

In short, any source capable of accelerating particles should be considered as a possible responsible of the $\gamma$-ray emission. Thus, given the great variety of objects that can emit at high energies and given the typical spatial extent and positional error of the Fermi sources, revealing the nature of unidentified $\gamma$-ray sources requires comprehensive multiwavelength studies both of the possible objects and their associated emission mechanisms.

4FGL J1846.9$-$0227 is a $\gamma$-ray source (first catalogued as 1FGL J1846.8-0233c) with no spatial extent detected, then classified as a point source. It was proposed to be associated with the MYSO candidate MSX6C G030.1981$-$00.1691 \citep[hereafter MSX G30;][]{munar2011}, based on the spatial correlation between both sources and considering the absence, according to the knowledge at that time, of other possible counterparts. Massive protostars have associated energetic jets and molecular outflows that can produce strong shocks during the interaction with the ambient gas. At these shocks, particles can be accelerated up to relativistic energies producing $\gamma$-ray emission, as some theoretical models predict \citep[e.g.,][]{araudo2012, bosch2010}.
 
A decade later, \citet{kerby21} analysed a sample of 174 unknown Fermi sources with a single X-ray/UV/optical counterpart. The authors developed a method to discern between pulsars and blazars, given that the most of the sources of 4FGL catalogue are these types of objects. In particular, the authors found two Swift-XRT X-ray sources as possible counterpart of the $\gamma$-ray source 4FGL J1846.9$-$0227.  The first one, SwXF4 J184650.7$-$022904, was rejected as possible pulsar or blazar because of its extreme photon index in the X-ray band ($\Gamma_{\rm X} \sim 7.9$). Moreover, the authors warned about the positional coincidence (shifted about 4\arcsec) of this Swift-XRT source with a dim star, suggesting that the high `blazar probability value' obtained should not be trusted, given that UV/optical emission would not be related to the blazar.  On the other hand, the second Swift source identified towards the region, SwXF4 J184651.6$-$022507, whose X-ray flux is three orders of magnitude smaller than the first one, was characterised as a likely blazar. 
 
It is worth mentioning that despite all the studies carried out to date, the exact origin of the $\gamma$-ray emission from 4FGL J1846.9$-$0227 remains unknown. In fact, in the last two incremental versions (4FGL-DR3 and 4FGL-DR4) of the fourth Fermi Large Area Telescope (LAT) catalogue of $\gamma$-ray sources (\citealt{abd22} and \citealt{ballet2023}, respectively),  this source still appears as `unknown'. However, it is important to mention that both catalogues refer to a possible association with the radio source TXS 1844$-$025 mentioned in \citet{honma2000}. Using the Japanese VLBI Network, the authors carried out radio continuum observations at 22 GHz towards a sample of 267 radio sources selected from existing radio surveys. In particular, they did not detect emission at 22 GHz towards the source TXS 1844$-$025. This radio source is in positional coincidence with the infrared source IRAS 18443$-$0231, which is catalogued as a planetary nebula (PN) candidate (\citealt{yang2016,irabor18}, and references therein).
 
Summarising, since the nature of 4FGL J1846.9$-$0227 is still unknown, we present a comprehensive and multi-wavelength analysis of all candidate sources in the region to identify definitely the origin of the $\gamma$-ray emission. It is important to highlight that this type of in-depth studies of specific sources are necessary to complement the statistical studies.

\section{Data}
\label{data}

The study of the origin of the $\gamma$-ray emission of 4FGL J1846.9$-$0227 involved the analysis of a multiwavelength set of data. In what follows, we describe each dataset.

\subsection{Radio continuum data} 

We used centimeter radio continuum data from several surveys:

\begin{itemize}
\item The continuum at 20\,cm ($\sim$1.4 GHz) was extracted from the Multi-Array Galactic Plane Imaging Survey \citep[MAGPIS;][]{hel06}, which has an angular resolution and sensitivity of about 5\s and 2~mJy\,beam$^{-1}$, respectively. Even with data from the D configuration, the resulting maps suffer missing flux from large-scale structure (above 1\m) to which the VLA is insensitive. To correct for this deficiency, the VLA images were combined with images from a 1400 MHz survey
made with the Effelsberg 100~m telescope \citep[][beam $\sim$ 9\farcm4]{reich1990}.

\item The continuum between between 4 and 8 GHz was extracted from GLOSTAR Galactic plane survey \citep{brunthaler21}, which uses the Very Large Array (VLA) with the D-array configuration. We retrieved the image averaged over the full bandwidth from the Image Server\footnote{Available at \url{https://glostar.mpifr-bonn.mpg.de/glostar/image_server}.}, which has a representative frequency of 5.8 GHz and are convolved with a $18^{\prime\prime}\times18^{\prime\prime}$ (FWHM) Gaussian kernel.  It is worth noting that these observations also are affected by the missing flux problem inherent to interferometric observations. For sources of a few arcmin, the non-recovered flux can be as large as 90\% (see Figure\,4 of \citealt{dokara21}). 

\item We also used data from the THOR Survey\footnote{https://www2.mpia-hd.mpg.de/thor/Overview.html}. This survey has mapped the Galactic plane between 1 and 2 GHz using the C-array configuration of the VLA at a native resolution of (FWHM) $\sim 12^{\prime\prime}\times14^{\prime\prime}$.

 \item The radio continuum at 3 GHz was extracted from the VLA Sky Survey \citep[VLASS;][]{lacy2020}. This survey covers the whole sky visible to the VLA (decl. $\geq -40 \deg$). The angular resolution and sensitivity are about 2\farcs5 and 100~$\mu$Jy\,beam$^{-1}$, respectively. The largest angular scale is about 30\arcsec.

\item The radio continuum at 5 GHz was extracted from the Co-Ordinated Radio 'N' Infrared Survey for High-mass star formation \citep[CORNISH;][]{purcell2008}. The survey was conducted using the Karl Jansky Very Large Array in B-configuration, yielding a 1\farcs5 resolution map with a noise level $\leq$ 0.4 mJy\,beam$^{-1}$. The largest angular scale is 15\arcsec.

\end{itemize}

\subsection{Mid-infrared and submillimeter data} 

\begin{itemize}

\item Mid-infrared data from the IRAC-Spitzer and MIPS-Spitzer at 8 and 24~$\mu$m were obtained from the GLIMPSE survey \citep{church09} and from the MIPSGAL survey \citep{carey05}, respectively. IRAC images have an angular resolution of about 1\farcs7 and the MIPSGAL angular resolution is 6\s at 24 $\mu$m.

\item Submillimeter data at 850 $\mu$m with a beam size of 13\farcs5 obtained from the Submillimetre Common-User Bolometer Array \citep[SCUBA;][]{difran2008} using the James Clerk Maxwell Telescope.

\item The $^{12}$CO J=3--2 data from the James Clerk Maxwell Telescope (JCMT) was extracted from the Canadian Astronomy Data Centre. The PI and proposal ID are Lumsden, S. and  M07AU08, respectively. The angular and velocity resolutions are 15$\arcsec$ and 0.4 km\,s$^{-1}$, respectively. The rms noise level is about 0.1 K.

\end{itemize}

\subsection{ALMA data} 

The data cube from the project 2015.1.01312 (PI: Fuller, G.;
Band 6) were obtained from the ALMA Science Archive\footnote{http://almascience.eso.org/aq/}. The single pointing observation for the target was carried out using the telescope configurations with L5BL/L80BL(m) 42/222 in the $12~\mathrm{m}$ array. The frequency range used in our analysis goes from 224.3 to 226.3~GHz. The angular resolution is 0\farcs7. The maximum recoverable scale is about 6\farcs2. The field of view is 25$\arcsec$. The frequency and velocity resolutions are 1.1~MHz and 1.4~\ks, respectively.
The continuum and line (10~\ks~averaged) sensitivity are 0.2 and 1.5~mJy~beam$^{-1}$, respectively. 

Although the data of the project passed the QA2 quality level, which assures a reliable calibration for ‘science ready’ data, we reprocessed the raw data using CASA 4.5.1 and 4.7.2 versions and the calibration pipeline scripts to check the final images. Particular care was taken with the different parameters of {\it clean} task. The images and spectra obtained from our data reprocessing —after several runs of the {\it clean} task, varying some of its parameters— were very similar to those obtained from the archival data. Briefly, the quality of the images automatically processed by the ALMA team could not be improved.

The task {\it imcontsub} in CASA was used to subtract the continuum from the spectral lines using a first-order polynomial. To that frequency ranges without molecular line emission were carefully selected. The continuum map at 240~GHz was corrected for primary beam. Several continuum subtraction tests were performed to ensure a reliable 240~GHz continuum map, which involved the selection of different free line regions of the spectrum. It is important to note that the beam size of the 240~GHz continuum data in the 12~m array provides a spatial resolution of about 0.02 pc ($\sim$ 4000 au) at 5.4 kpc, which is the kinematic distance assigned to the MYSO candidate MSX G030.

\subsection{XMM-Newton data}

The positional error ellipse of 4FGL J1846.9$-$0227 falls within the field-of-view of the XMM-Newton observations 0651680301 and 0823990801, hereafter Obs1 and Obs2, respectively. The observations were pointed to the supergiant fast-X-ray transient IGR J18462$-$0223, and the centroid of the Fermi source is located $\sim12^{\prime}$ off-set from the centre of the X-ray cameras. The duration of Obs1 and Obs2 are 33.4 and 35.7~ks, respectively. We used the Science Analysis System (SAS, version 21.0.0) and HEASoft (version 6.33.2) for data reduction and spectral analysis. We applied the standard procedure to obtain a calibrated event file for each camera of each observation using the updated set of calibration files (CCF). We use these event files to produce light curves in the high-energy domain ($>10~\mathrm{keV}$) and filtered out periods of high-count rate associated with background flares. We applied additional filtering to include events with FLAG = 0 and PATTERN $\leq$ 12 and 4 for EPIC-pn and EPIC-mos cameras, respectively. In Table \ref{table_XMM} we report a summary of the X-ray observations.

\begin{table}
\centering
\caption{Summary of the XMM-Newton observations. The {\it Net exposure} column reports the effective exposure time after filtering the periods of high-count rate.}
\begin{tabular}{cccc}
    \hline\hline
         Obs. ID (date)&  Camera &Science mode & Net exposure (ks)\\
         \hline
         0651680301      & EPIC-mos1 & Full Frame   & 30.8 \\ 
         (Apr. 18, 2011) & EPIC-mos2 & Full Frame   & 31.6 \\
                         & EPIC-pn   & Full Frame   & 15.1\\
         \hline
         0823390801      & EPIC-mos1 & Small Window & 31.3 \\
         (Oct. 22, 2018) & EPIC-mos2 & Timing       & 31.1 \\
                         & EPIC-pn   & Full Frame   & 14.5 \\
        \hline
\end{tabular}
\label{table_XMM}
\end{table}

\section{Results}
\label{results}

\begin{figure}
\centering
\includegraphics[width=8.5cm]{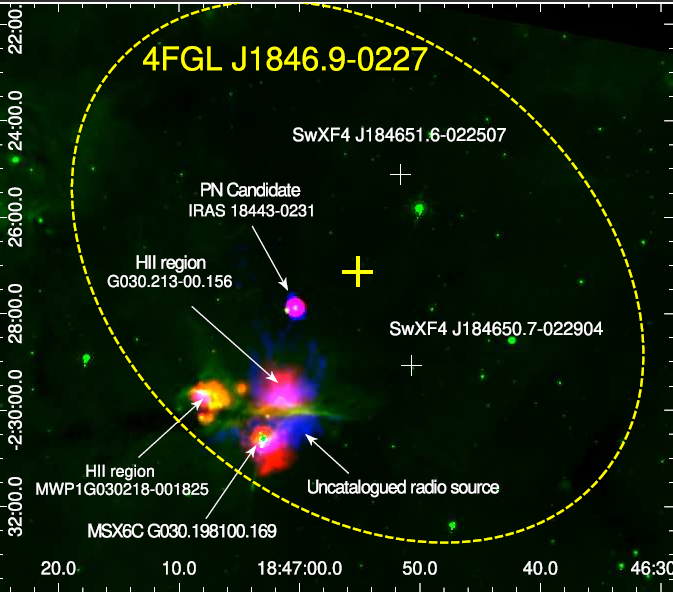}
\smallskip
\caption{Large scale three colour image of the 95\% positional error confidence region associated with the Fermi-LAT source 4FGL J1846.9$-$0227 (dashed-ellipse). The yellow cross indicates the central position of the     $\gamma$-ray source. The Spitzer infrared emission at 8 and 24 $\mu$m are shown in green and red, respectively, and the MAGPIS radio continuum emission at 20~cm is shown in blue. The main sources in the region are labelled, some of them are candidates for being the counterpart of the Fermi source. }
\label{present1}
\end{figure}

Fermi-LAT sources typically have positional errors of a few arcminutes,  which implies that there are often several candidates to be the counterpart of the high-energy source.
Figure\,\ref{present1} shows a three colour image with 8 and 24 $\mu$m, and 20 cm emissions represented in green, red, and blue, respectively. The centre and the 95\% positional error confidence ellipse of the Fermi source 4FGL J1846.9$-$0227 are represented by the yellow cross and the dashed-yellow ellipse, respectively.  We carried out the search for the Fermi counterpart in the region within such an ellipse.

\begin{figure*}
\centering
\includegraphics[width=18cm]{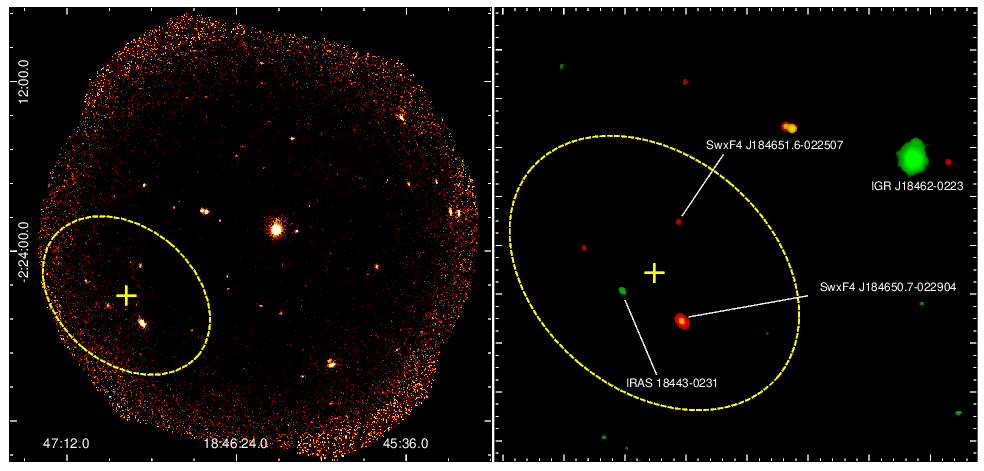}
\smallskip
\caption{Left: Colour-scale shows the whole field of view of the X-ray emission between 0.5 and 8.0 keV from XMM-Newton.  Right: Two-colour image of the X-ray emission between 0.5 and 2.0 keV (red) and 2.0 and 8.0 keV (green). It is marked the position of IGR J18462$-$0223 (the original target of the observations), the position of the Swift sources studied by \citet{kerby21}, whose nature is discussed in this work, and the position of IRAS 18443-0231 which presents considerable hard X-ray emission. The not marked `soft source' within the ellipse, the red one towards the northeast, is related to the HD 173763 star, which is located at a distance of about 330 pc, and it is not considered for our analysis. In both panels, the yellow cross and ellipse indicate the best-fit position and positional error of the Fermi source, respectively.}
\label{present1X}
\end{figure*}

Figure\,\ref{present1X}-left panel shows the whole field of view of the XMM-Newton observations in the X-ray energy band between 0.5 keV and 8.0 keV. This image was obtained by combining the 6 images and exposure maps of the EPIC-mos1, EPIC-mos2 and EPIC-pn cameras of each observation using the {\it emosaic} task of SAS. We applied a Gaussian smoothing to increase the S/N ratio. Figure\,\ref{present1X}-right panel shows a two-colour image of the X-ray soft/medium and hard emissions between 0.5 and 2.0 keV (red) and 2.0 and 8.0 keV (green), respectively. No hint of diffuse emission coincident with the Fermi source and some point sources can be clearly observed. The brightest X-ray point source, by far, coincident with the Fermi positional error ellipse is 4XMM J184650.6$-$022907 (associated with SwXF4 J184650.7$-$022904), a soft source with no counts detected above 3.0 keV. Additionally are indicated the soft X-ray source 4XMM J184651.2$-$022504 (associated with SwXF4 J184651.6$-$022507) and the hard X-ray source 4XMM J184700.4$-$022752 (associated with the PN candidate IRAS 18443$-$0231). These sources are analysed below.

\begin{figure*}
\centering
\includegraphics[width=17cm]{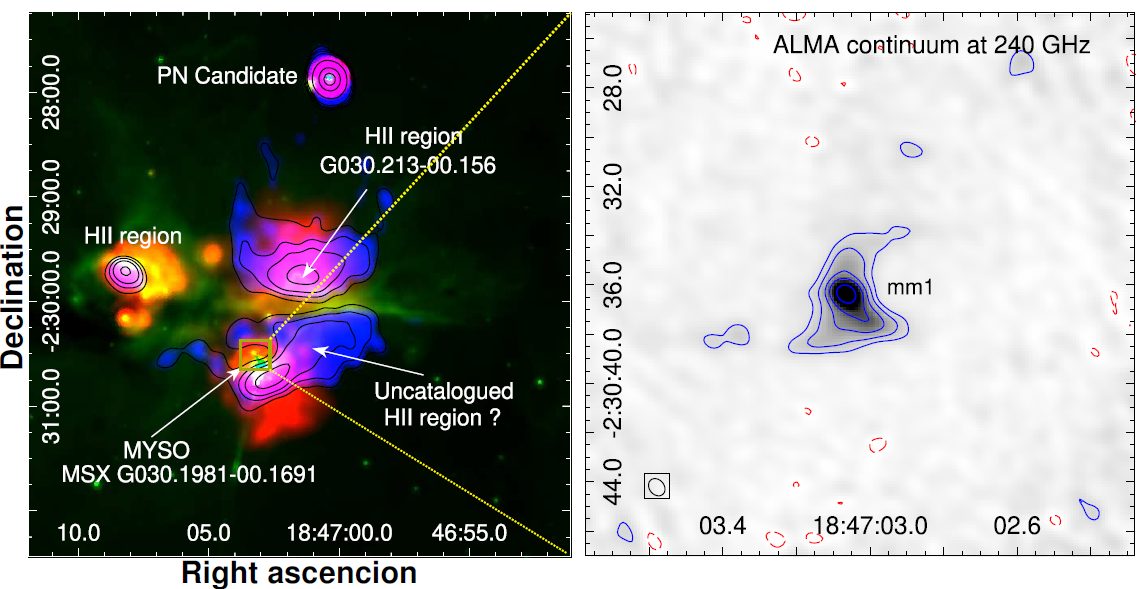}
\caption{Left panel: Close-up view of Fig.\,\ref{present1} towards the \hii~regions area. The Spitzer infrared emission at 8 and 24 $\mu$m are shown in green and red, respectively, and the MAGPIS radio continuum emission at 20~cm is shown in blue. The black contours highlight the radio continuum emission at 20~cm. Levels are at 3.2, 3.3, 3.4, and 3.5 mJy beam$^{-1}$. Right panel: Grey-scale and blue contours represent the ALMA continuum emission at 1.3~mm. Levels are at 0.5, 2, 5, 8, and 10 mJy beam$^{-1}$. The dashed red contours represent the 3~$\sigma$ negative continuum emission.}
\label{alma-continuum}
\end{figure*}

The region considered for the analysis includes four centimeter radio continuum sources (see Fig.\,\ref{present1} and Fig.\,\ref{alma-continuum}-left). G030.213$-$00.156 (hereafter G30.21) is catalogued as an \hii~region  with a radio recombination line at 100~\ks, which corresponds to the near kinematic distance of about 5.4~kpc \citep{anderson15}. Southwards G30.21, we identify an uncatalogued extended radio continuum source. It is worth noting that both radio sources, which seem not to be connected, exhibit an evident flattening facing a conspicuous dust filament seen at 8 and 24 $\mu$m (see Fig.\,\ref{present1} and Fig.\,\ref{alma-continuum}-left). It is possible that both radio sources compress such a dust filament generating its flattened morphology. It is important to mention that \citet{anderson15} used Green Bank Telescope radio observations, whose angular resolution of about 87$\arcsec$ prevented them from resolving both radio sources, which were considered as a single one. 

Interestingly, the peak of emission of the \hii~region G30.21 is in positional coincidence with an illuminated protrusion of the filament, while the peak of emission of the new radio continuum source appears located close to the MYSO candidate MSX G30, which seems to be surrounded by diffuse and extended gas (see Fig.\ref{alma-continuum}-left). \citet{anderson2009} determined a kinematic distance of about 5~kpc for MSX G30. The morphological coincidence between the uncatalogued radio continuum source and the dust filament and the MYSO suggests that this new radio source would be located at the same distance of about 5~kpc of the complex. Both radio sources have 24~$\mu$m emission in coincidence with their radio continuum peaks, suggesting that the origin of the uncatalogued source is thermal. However we can not discard that this new radio source is a SNR, which would be of great importance for the study of the origin of the $\gamma$-ray emission, hence, we investigate such a source through an spectral index analysis (see Sect.\,\ref{index}).

On the other hand, it can also be appreciated two additional compact radio sources in the region, IRAS 18443$-$0231 and MWP1G030218$-$001825, which have been catalogued as a possible planetary nebula, and a compact \hii~region \citep{anderson11}, respectively. Given that the nature of IRAS 18443$-$0231 is still an open issue and it is associated with the X-ray source 4XMM J184700.4$-$022752, it deserves more analysis in the context of this work (see Sect.\,\ref{pnsect}). In such a context, the three X-ray sources indicated in Fig.\,\ref{present1X}-right panel are analysed in detail.

In addition, the possible contribution to the $\gamma$-ray emission of the MYSO candidate MSX G30 (see Fig. \ref{alma-continuum}), which is embedded in the dust clump AGAL G030.1978$-$0.1683, is analysed (see Sect.\,\ref{ysoanalysis}).

\begin{figure*}
\centering
\includegraphics[width=16cm]{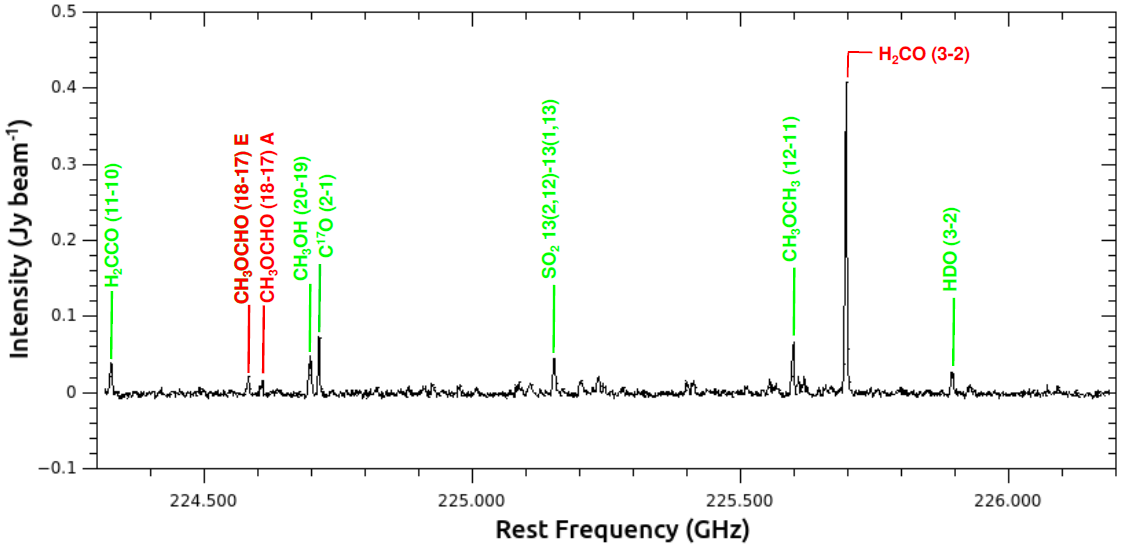}
\caption{ALMA Band 6 beam averaged spectrum towards the core mm1. The zero moment maps of the red labelled molecules are exhibited in Fig.\ref{zero-moments}.}
\label{spw3}
\end{figure*}

\begin{figure*}
\centering
\includegraphics[width=17cm]{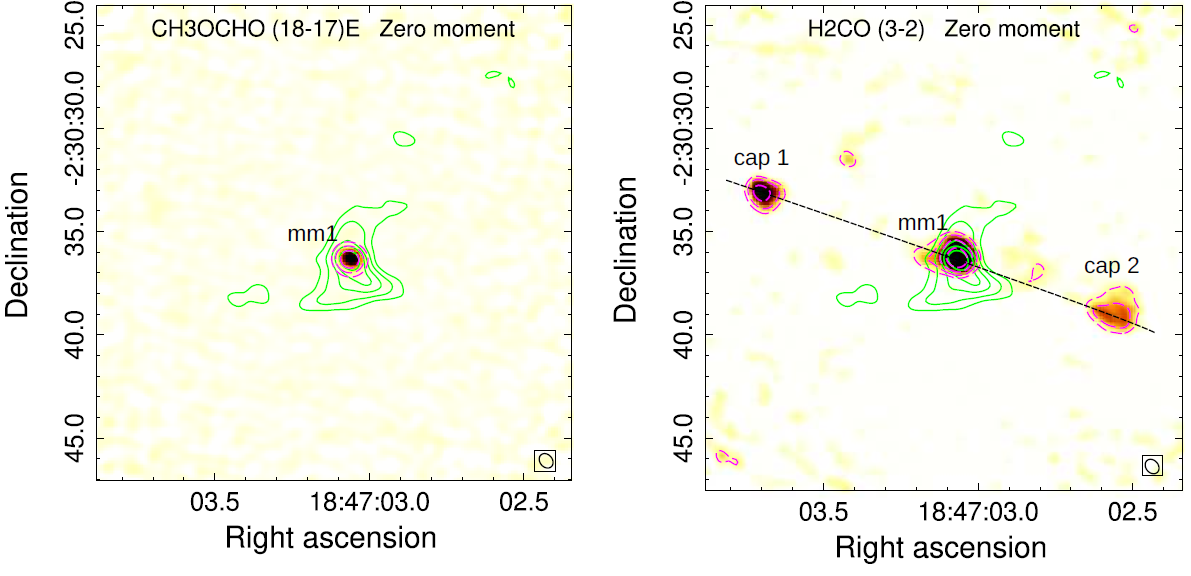}
\caption{Colour-scale and dashed magenta contours show the integrated emissions of the CH$_3$OCHO (18--17)E (left panel) and H$_2$CO J=3--2 (right panel) transitions. Levels are at 0.02, 0.04, and 0.07 Jy\,beam$^{-1}$ and at 0.2, 0.5, 1, and 2 Jy\,beam$^{-1}$ for CH$_3$OCHO and H$_2$CO, respectively. The green contours represent the ALMA continuum emission at 1.3~mm. Levels are at 0.5, 2, 5, 8, and 10 mJy\,beam$^{-1}$. The beam of the observation is indicated at the right bottom corner.}
\label{zero-moments}
\end{figure*}

\subsection{The nature of the uncatalogued radio source}
\label{index}

In this subsection, we analyse the possibility that the new radio source is a SNR, for which a radio spectral index study is carried out. We use the averaged image between 1.4 and 1.8 GHz from the THOR survey\footnote{Retrieved from \url{https://thorserver.mpia.de/data/continuum/}.}.    
This image was convolved with a $18^{\prime\prime}\times18^{\prime\prime}$ (FWHM) Gaussian kernel to match the resolution of the GLOSTAR image. 

We obtained the radio continuum fluxes of the \hii~region G30.21 and the new radio source by integrating the emission over elliptical regions that approximately match the $3.4~\mathrm{mJy\,beam^{-1}}$ contour level indicated in Fig.\,\ref{alma-continuum}-left panel. For the \hii~region G30.21, we obtained flux densities $S_{\nu}$ of $47 \pm 2$  and $53 \pm 1$ mJy at 1.6 (THOR) and 5.8 (GLOSTAR) GHz, respectively. For the southern region, we obtained flux densities of $81 \pm 2$ and $69 \pm 1$ mJy at 1.6 and 5.8 GHz, respectively. The mean noise (rms) at both frequencies is $\sim 1~\mathrm{mJy\,beam^{-1}}$ and it was measured over regions free of source emission.

From the flux densities derived above, we calculated radio spectral indexes $\alpha$ ($S_{\nu}\propto \nu^{\alpha}$). We obtained $\alpha = 0.10 \pm 0.04$ and $-0.12 \pm 0.03$ for the \hii~region G30.21 and the southern uncatalogued region, respectively.  Our results confirm the thermal nature of the radio emission from \hii~region G30.21, and strongly suggest that the new radio continuum source is another \hii~region, named G030.195$-$00.168 following the naming criteria of \citet{anderson15}. It is important to mention that this result is consistent with the presence of 24 $\mu$m emission in positional coincidence with the 1.4 GHz radio continuum emission. Therefore, it seems reasonable to rule out this new radio source as a possible cause of the $\gamma$-ray emission.

\subsection{The MYSO candidate MSX G30}
\label{ysoanalysis}

The MYSO candidate MSX G30 is catalogued as a massive protostar candidate with a bolometric luminosity of about $3 \times 10^4$~L$_{\odot}$ \citep{lumsden2013}. It is well known that MYSOs have associated massive bipolar molecular outflows which can produce strong shocks when they interact with the surrounding medium. In these conditions particle acceleration at relativistic velocities can occur leading to $\gamma$-ray emission, as some theoretical models predict \citep[][and references therein]{bosch2010}. 

Crossing the Fermi First Year Catalogue with some catalogues of MYSO candidates, \citet{munar2011} obtained a list of MYSOs that are spatially coincident with Fermi sources, and therefore, potentially responsible for the $\gamma$-ray emission.  In particular, the authors found a spatial correlation between the Fermi source 1FGL J1846.8$-$0233c (a.k.a. 4FGL J1846.9$-$0227) and MSX G30. Therefore, we searched in the literature for evidence of molecular outflow activity associated with this MYSO candidate. 

Using intermediate angular resolution data (about 15$\arcsec$) from the $^{13}$CO/C$^{18}$O (3--2) Heterodyne Inner Milky Way Plane Survey \citep[CHIMPS;][]{rigby2016}, \citet{yang2018} searched for massive outflows associated with a sample of 919 ATLASGAL clumps. In particular, the authors did not find any evidence of outflow activity towards the source AGAL G030.1978$-$0.1683, which is the dust condensation in which MSX G30 is embedded. On the other hand, we searched for Extended Green Objects (EGOs) in the catalogue of \citet{cyga2008}. EGOs are usually associated with massive molecular outflow activity. However, there is not any EGOs catalogued in the region. 

In this context, we analysed high resolution and sensitivity continuum and line ALMA data at Band 6 searching for traces of molecular outflow activity at core scale.

Figure\,\ref{alma-continuum}-right panel shows a close-up view of the MSX G30 region. The gray-scale and blue contours show the ALMA continuum emission at 1.3 mm. It can be seen the presence of a single dust core surrounded by fainter extended emission. The lack of evidence of fragmentation suggests that the bolometric luminosity of about $10^4$~L$_{\odot}$ is associated with this single source, which confirms that MSX G30 is indeed a massive protostar.

Figure\,\ref{spw3} shows a beam average spectrum ranging from about 224.3 to 226.3~GHz (Band 6) towards the core mm1. Several molecular lines have been detected in this spectral window, but for the purpose of our analysis we will focus on two of them: CH$_3$OCHO (18,17) and H$_{2}$CO (3--2). 
Figure\,\ref{zero-moments}-left panel shows the integrated emission map of CH$_3$OCHO (18,17) E transition. The spatial distribution of this molecule  appears well localised onto the position of the core mm1. The mere presence of this transition, whose typical desorption temperature is about 120 K (e.g. \citealt{busch2022}), shows that mm1 is a hot molecular core, which confirms that active star formation is ongoing inside.

Figure\,\ref{zero-moments}-right panel shows the integrated emission map of H$_{2}$CO J=3--2 transition. It can be seen that the spatial distribution of this molecule exhibits a main peak which perfectly match the position of the core mm1. In addition,  it can be noticed two emission features with no associated continuum emission, which are perfectly aligned with the core mm1 (see dotted black line in Fig.\,\ref{zero-moments}-right). Taking into account that this molecule usually traces dense gas and molecular outflows \citep{tycho2021, okoda2020}, it is very likely that both structures outside the core position are showing signs of outflow activity. In particular, both structures could be tracing the densest gas of the molecular outflows, likely the caps. Given that the velocity difference between both caps are about 3~\ks, we can infer that the lobes are extended mostly at the plane of the sky. The lineal distance between both caps are about 0.5~pc, which is in agreement with typical outflows sizes. 

\subsection{Source SwXF4 J184651.6$-$022507}

We note that 4XMM J184651.2$-$022504 is the XMM-Newton counterpart of the Swift source SwXF4 J184651.6$-$022507 mentioned in Sect.\,\ref{intro} (see Fig.\,\ref{present1X}). According to \citet{kerby21}, this source has
an X-ray flux three orders of magnitude smaller than SwF4 J184650.7$-$022904 (see Sect.\,\ref{xraysect}) and it was characterised as a likely blazar. 

This faint X-ray source is related to the mid-IR source WISE J184651.22$-$022505.5 from the AllWISE data release Catalogue of \citet{cutri13}. The reported magnitudes for this source in the four mid-IR observed bands are: [3.4]=10.50, [4.6]=10.46, [12.0]=10.08, and [22.0]=5.09 (quality flags AAUB). Even though the quality at the 12 $\mu$m band indicates that it is an upper limit, it can be used to locate this source in a colour-colour ([3.4]-[4.6] vs. [4.6]-[12.0]) diagram. This kind of diagrams are very frequently used to determine a blazar association (e.g. \citealt{dabrusco12}). The source lies very far from the blazars region in such a diagram, in fact, it is located in the region in which the stars are expected to be. Moreover, giving that the source lies very close to the galactic plane (l = $-$0.08 deg),  a galactic origin is more likely.  Hence, we discard the blazar nature and its responsibility for the high-energy emission.
\\

\subsection{Source SwXF4 J184650.7$-$022904}
\label{xraysect}

Source 4XMM J184650.6$-$022907 is the XMM-Newton counterpart of the Swift source SwXF4 J184650.7$-$022904 (see Fig.\,\ref{present1X}).
According to the XMM-Newton Serendipitous Source Catalogue, Thirteenth Data Release\footnote{Catalogue available through the XMM-Newton Science Archive interface at \url{https://www.cosmos.esa.int/web/xmm-newton/xsa}} (4XMM-DR13), this source presents significant variability: the EPIC count-rate between 0.2 and 12 keV increases from $(9.0\pm0.4) \times 10^{-2}~\mathrm{cts\,s^{-1}}$ for Obs1 (2011) to $(17.4\pm0.6) \times 10^{-2}~\mathrm{cts\,s^{-1}}$ for observation Obs2 (2018). 

\begin{figure}
\centering
\includegraphics[width=8cm]{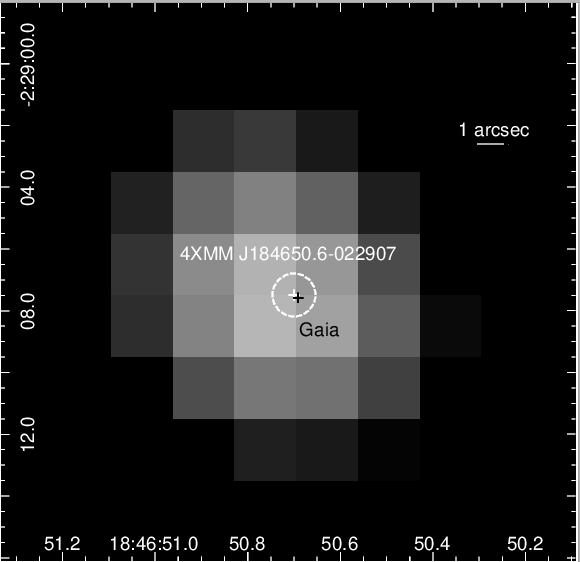}
\caption{Zoom-in of the exposure-corrected image of  Fig.\,\ref{present1X} (right). The white cross and circle are the position and positional error of 4XMM J184560.6$-$022907 reported in the 4XMM-DR13 catalogue, respectively. The position of the spectroscopic binary star BD$-$02\,4739 from Gaia DR3 is marked with a black cross. }
\label{x+bd}
\end{figure}

Looking for possible counterparts of 4XMM J184650.6$-$022907, we found that it is related to mid-IR source 
WISE J184650.7-022904 \citep{cutri13} with magnitudes [3.4]=8.861, [4.6]=6.951, [12.0]=7.108, and [22.0]=8.106 (quality flags AAAU). Again, from a colour-colour ([3.4]-[4.6] vs. [4.6]-[12.0]) diagram, this source is not compatible with a blazar.  
We also found that the star BD$-$02\,4739 (at 18:46:50.69 $-$02:29:07.30, J2000), catalogued as a spectroscopic binary in the Non-single stars Catalogue by GAIA DR3 \citep{gaia22}, is considered the optical counterpart in the 4XMM-DR13 catalogue. Indeed, it lies at the same coordinate of 4XMM J184650.6$-$022907 within a positional difference less that 0.1 arcsec (see Fig.\,\ref{x+bd}). Taking into account this perfect positional coincidence, it is very likely that both sources are related, and hence, it deserves more analysis in the context of this work.  

\subsubsection{Spectroscopic binary star BD$-$02\,4739}
\label{SED}

As noted in previous section, it is very likely that the spectroscopic binary star BD$-$02 4739 is the optical counterpart of 4XMM J184650.6$-$022907. The distance and orbital period reported by GAIA DR3 are $452\pm12~\mathrm{pc}$ and $53~\mathrm{d}$, respectively \citep{gaia22}. Several other optical/ir associations are reported in the 4XMM-DR13 catalogue and Simbad for this star, which are useful to determine its spectral type through a spectral energy distribution (SED) analysis. 

In a binary system, in principle both components contribute to the fluxes observed at different wavelengths, hence a SED study must be done with care. In this case, we conducted a SED analysis based on the statement that one of the components of this binary system should be a black hole, a neutron star or a white dwarf, issue discussed in Sect.\,\ref{discussion}, and hence we assume that the analysed fluxes for the SED came mainly from a donor stellar companion.

In Table\,\ref{star_flux} we report the associated sources from the WISE Source Catalogue (AllWISE, the flux at band W4 is reported as an upper limit), the Galactic Legacy Infrared Mid-Plane Survey Extraordinaire (GLIMPSE) performed with the Spitzer Space Telescope, the Two Micron All-Sky Survey (2MASS), the GAIA Data Release 3 (Gaia DR3), the Tycho 2 Main Catalogue (Tycho-2), and the XMM-Newton Optical Monitor (XMM-OM) Source Catalogue (XMM-SUSS5.0). We have excluded the optical source 1237668687974826071 from the Sloan Digital Sky Survey Data Release 12 (SDSS DR12) because several photometric flags are reported. Flux densities ($f_\lambda$) were calculated from the magnitudes $m$ reported in the catalogues as $f_\lambda= f_0 10^{-0.4m}$, where $f_0$ is the zero-point flux, whose values were obtained from \citet{Rodrigo20}. The fluxes were corrected for extinction, which was obtained from the Gaia-2MASS 3D maps of Galactic interstellar dust of \citet{lallement19}. In the line of sight of BD$-$02\,4739 ($l = 30^{\circ}.20$, $b = -0^{\circ}.11$), we obtained a visual extinction $A_V = 1.23$ for a distance of $450~\mathrm{pc}$\footnote{Extinction calculator available at: \url{https://astro.acri-st.fr/gaia_dev/}. $A_V=1.23$ corresponds to a reddening $E(B-V) = A_V/R_V = 0.4$ using the standard $R_V=3.1$.}. In Table\,\ref{star_flux} we report the dereddened fluxes ($f_\lambda^*$), calculated using the VO SED Analyzer (VOSA\footnote{Available at: \url{http://svo2.cab.inta-csic.es/theory/vosa/}}, \citealt{bayo08}).

\begin{table*}
\small
\centering
\caption{UV, optical and infrared fluxes of BD-02 4739/4XMM J184650.6$-$022907. $f_\lambda$ is the observed flux and $f_\lambda^{*}$ the corresponding dereddened flux considering the extinction $A_V = 1.23$.}
\small
\begin{tabular}{cccc}
    \hline\hline
Survey / Source ID &  Filter / $\lambda$ [A]  & $f{_\lambda}$ [$\mathrm{erg\,s^{-1}\,cm^{-2}\,A^{-1}}]$ & $f_\lambda^*$ [$\mathrm{erg\,s^{-1}\,cm^{-2}\,A^{-1}}]$\\
         \hline 
AllWISE /            & W1 / 33526   & $1.42\pm0.04 \times 10^{-14}$ & $1.54\pm0.04 \times 10^{-14}$ \\
184650.68$-$022907.4 & W2 / 46028    &  $4.03\pm0.08 \times 10^{-15}$ & $4.27\pm0.08 \times 10^{-15}$\\
                     & W3 /  115608      & $1.01\pm0.06 \times 10^{-16}$ & $1.05\pm0.06\times 10^{-16}$  \\
                     & W4 /  220833      & $<2.09\times 10^{-18}$ & $<2.97\times 10^{-18}$  \\
 \hline
Spitzer GLIMPSE /   &   IRAC 3.6 / 35074  & $1.01\pm0.02 \times 10^{-14}$ & $1.09\pm0.02 \times 10^{-14}$ \\
G030.1965$-$00.111  &   IRAC 4.5 / 44365   &$4.19\pm0.17 \times 10^{-15}$ & $4.44\pm0.18 \times 10^{-15}$ \\
                    &   IRAC 5.8  / 56280  & $1.87\pm0.05 \times 10^{-15}$ & $1.96\pm0.05 \times 10^{-15}$\\
                    &    IRAC 8.0 / 75890  & $5.55\pm0.15 \times 10^{-16}$ & $5.77\pm0.16 \times 10^{-16}$ \\
\hline
2MASS /              &   J / 12350  & $2.30\pm0.06 \times 10^{-13}$ & $3.25\pm0.08 \times 10^{-13}$ \\
18465069$-$0229072   &   H / 16620   & $1.47\pm0.04 \times 10^{-13}$ & $1.83\pm0.04 \times 10^{-13}$\\
                          &   Ks / 21590   & $6.56\pm0.10 \times 10^{-14}$ & $7.56\pm0.12 \times 10^{-14}$\\ 
\hline
Gaia DR3 /              &   BP / 5035  & $2.12\pm0.01 \times 10^{-13}$ & $7.41\pm0.03 \times 10^{-13}$ \\
4259072330161989504     &   G / 5822   & $2.73\pm0.01 \times 10^{-13}$ & $7.31\pm0.02 \times 10^{-13}$\\
                        &   RP / 7619   & $3.17\pm0.01 \times 10^{-13}$ & $6.52\pm0.03 \times 10^{-13}$\\
\hline     
Tycho-2  /              &   B / 4280  & $1.11\pm0.10 \times 10^{-13}$ & $5.25\pm0.47 \times 10^{-13}$ \\
TYC2 5118-160-1       &   V / 5340   & $2.32\pm0.10 \times 10^{-13}$ & $7.70\pm0.34 \times 10^{-13}$\\                     
\hline
XMM-OM /              &   UVM2 / 2326 &$1.24\pm0.03 \times 10^{-15}$ & $2.88\pm0.08 \times 10^{-14}$\\
50813589     &             &  \\
\hline
\end{tabular}
\label{star_flux}
\end{table*}

\begin{figure}
\centering
\includegraphics[width=9cm]{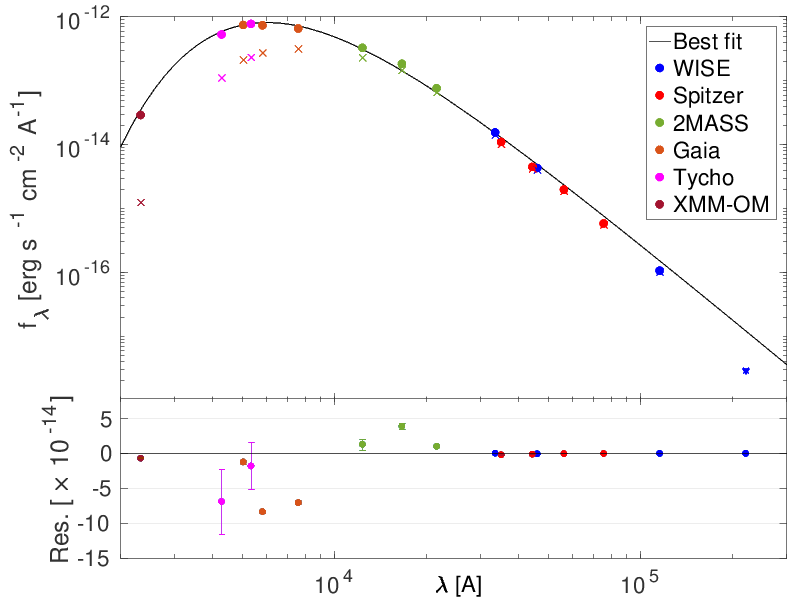}
\caption{UV, optical and infrared observed fluxes $f_\lambda$ (crosses) and dereddened fluxes $f_\lambda^{*}$ (filled circles). The black line is the best fit black-body. The error bars are only shown in the residuals plot.}
\label{sed}
\end{figure}

We used VOSA to fit the deredenned fluxes with a black-body function. The fit parameters are the effective temperature $T_{eff}$ and the dimensionless normalization $A$:
\begin{eqnarray}
{f_{\lambda}^{*} = A \frac{2hc^2}{\lambda^5}\frac{1}{\mathrm{exp}(hc/\lambda k T_{eff})-1}},
\label{SED_star1}
\end{eqnarray}
where $\lambda$ is the wavelength, $c$ the light speed, $k$ the Boltzmann constant and $h$ the Planck constant. 
The normalization $A$ is related with the angular size of the source,
\begin{eqnarray}
{A = \left( \frac{R}{D} \right)^2},
\label{SED_star2}
\end{eqnarray}
where $R$ is the stellar radius and $D$ the distance. 
In Fig.\,\ref{sed} we show the fluxes of Table\,\ref{star_flux} and the result of the fit. We obtain $T_{eff}=4800 \pm 30~\mathrm{K}$ (error correspond to 90\% confident level). The stellar flux is $F = 7.5\pm0.1\times10^{-9}~\mathrm{erg\,s^{-1}\,cm^{-2}}$ and the bolometric luminosity is $L = 48\pm1~\mathrm{L_\odot}$ for a distance of $452~\mathrm{pc}$. 
Using the best-fit value of $A=7.75 \times 10^{-19}$, we obtain a stellar radius $\sim 17~\mathrm{R_\odot}$.

\begin{figure}
\centering
\includegraphics[width=9cm]{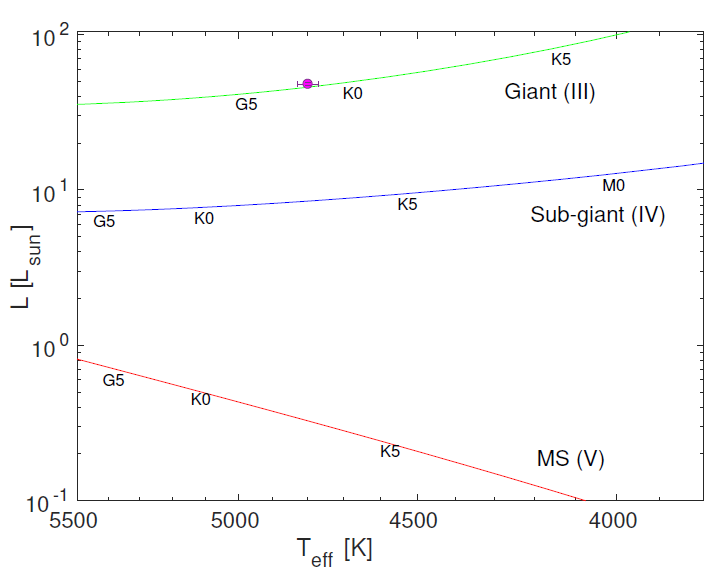}
\caption{HR diagram with the main sequence (MS), sub-giant, and giant luminosity classes. Some spectral types have been included as a reference. The point corresponds to the location of BD$-$02 4739 in the diagram.}
\label{fig_HR}
\end{figure}

To determine the spectral type of BD$-$02 4739, we constructed the HR diagram shown in Fig. \ref{fig_HR}, where we included the location of the main sequence, sub-giant, and giant stars, corresponding to luminosity classes V, IV and, III, respectively (taken from \citealt{deJager87}).
Hence, we suggest that the donor star, i.e. the member that would contribute the most to the IR and optical fluxes, should be a low-mass giant star.

\subsubsection{X-ray spectral analysis}
\label{spec_X}

We extracted the X-ray spectra of 4XMM J184650.6$-$022907 from the 3 cameras of each XMM-Newton observation, yielding to a set of 6 spectra. The source extraction region was a circle of $24^{\prime\prime}$~in radius and background spectra were extracted from nearby circular regions in the same CCD as the source region. 
We use the {\it evselect} task of SAS with a minimum of 15 counts per bin. 
We obtained 209 (EPIC-mos1), 245 (EPIC-mos2), and 416 (EPIC-pn) counts for Obs1, and 359 (EPIC-mos1), 391 (EPIC-mos2), and 415 (EPIC-pn) counts for Obs2, in the $0.5-3.0~\mathrm{keV}$ energy band. The spectra were displayed and analysed with the Xspec software (version 12.14.0), which is included in the HEASoft package. 

\begin{table*}
\centering
\caption{Best fit parameters for the optically thin plasma ($tbabs \times apec$) for 4XMM J184650.6-022907. The ``='' symbol indicates that the corresponding parameter for Obs2 is set equal to Obs1 during the fitting procedure. $Abund.$ is the metal abundance relative to the Sun and $\eta$ is the normalisation, defined as the emission measured of the gas scaled by the distance: $\eta = 10^{-14}\int{n_e n_H dV /(4\pi D^2)}$, where electron and proton densities are expressed in cm$^{-3}$, the gas volume element $dV$ in cm$^{3}$, and the source distance $D$ in cm. $F_X$ and $L_X$ are the flux and luminosity corrected for absorption in the $0.5-3.0~\mathrm{keV}$ energy band, respectively. $L_X$ is calculated for a distance of $452~\mathrm{pc}$.}
\begin{tabular}{c|cc|cc}
\hline
\hline
Parameter               &  Obs1 & Obs2            & Obs1 & Obs2 \\
\hline
$\chi^2_r$              &  \multicolumn{2}{c|}{1.27} & \multicolumn{2}{c}{1.04} \\
$N(\mathrm{H})$ ($\times 10^{22}~\mathrm{cm^{-2}})$ & $0.28^{+0.13}_{-0.05}$ & = & $0.32^{+0.08}_{-0.06}$ &  =   \\
$kT$ (keV)    & $0.91^{+0.06}_{-0.12}$ & = & $0.79 ^{+0.06}_{-0.06}$  & $0.96 ^{+0.07}_{-0.09}$ \\
$Abund.$        &  $0.12^{+0.04}_{-0.03}$ & = & $0.19 ^{+0.15}_{-0.07} $ & $0.11^{+0.05}_{-0.04}$ \\
$\eta$ ($\times 10^{-4}$) & $ 2.26^{+0.80}_{-0.39}$ & $ 4.27^{+1.20}_{-0.75}$ & $2.04 ^{+0.67}_{-0.58} $ & $4.92 ^{+1.30}_{-0.92}$ \\
$F_X$ ($\times 10^{-13}~\mathrm{erg\,s^{-1}\,cm^{-2}}$) & $ 1.52^{+0.59}_{-0.21} $ & $ 2.87^{+1.08}_{-0.39}$ & $1.72^{+0.41}_{-0.28}$  & $3.13^{+0.61}_{-0.45}$ \\
$L_X$ ($\times 10^{30}~\mathrm{erg\,s^{-1}}$)   &   $ 3.72 ^{+1.44}_{-0.51} $ & $ 7.01^{+0.95}_{-2.64}$ & $4.21 ^{+0.69}_{-0.99}$  & $7.64 ^{+1.09}_{-1.51}$ \\
\hline 
\hline
\end{tabular}
\label{tabla_apec}
\end{table*}

\begin{figure}
\centering
\includegraphics[width=6.3cm, angle=-90]{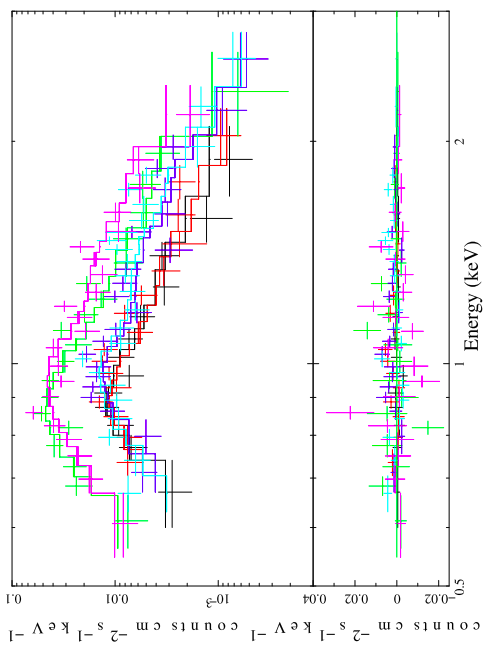}
\caption{X-ray spectra of the point source 4XMM J184650.6$-$022907 between 0.5 and 3.0 keV, for XMM-Newton Obs1 (EPIC-mos1: black, EPIC-mos2: red, EPIC-pn: green) and Obs2 (EPIC-mos1: blue, EPIC-mos2: cyan, EPIC-pn: magenta). The solid lines correspond to the best fit absorbed thermal plasma ($tbabs \times apec$) with N$_H$ tied together between observations and all the remaining parameters let to vary free. }
\label{figXspec}
\end{figure}

We visually inspected the 6 spectra and found X-ray emission coming from a soft and featureless continuum, without evidence of spectral lines (Fig. \ref{figXspec}).
We first fit each observation separately with different emission models in the $0.5-3.0~\mathrm{keV}$ energy band to look for possible variations of model parameters between Obs1 and Obs2. 
To explore the non-thermal scenario we use an absorbed power law ($tbabs \times powerlaw$). We obtain an unrealistic soft photon index ($\Gamma_X \sim 8$ and $\sim 5$ for Obs1 and Obs2, respectively) and a quite poor fit ($\chi^2_r \sim 1.4$ and 1.8, for Obs1 and Obs2)\footnote{To characterise the goodness of the fit, we report the reduced $\chi^2$, i.e. $\chi^2_r= \chi^2 / \mathrm{d.o.f.}$, where $d.o.f.$ are the degrees of freedom.}.
To explore the thermal nature of the source, we fitted the spectra with a blackbody, a bremsstrahlung, and an optically thin plasma. We obtained poor results for the blackbody ($\chi^2_r \sim 1.6$ and 2.0, for Obs1 and Obs2) and bremsstrahlung ($\chi^2_r \sim 2$ for Obs1 and Obs2), and an acceptable fit for the thin plasma ($tbabs \times apec$).
For this model, the 90\% confidence range for the hydrogen column densities are $N(\mathrm{H}) \sim (0.23 - 0.53)\times 10^{22}~\mathrm{cm^{-2}}$ (Obs1)  and $\sim (0.24 - 0.40)\times 10^{22}~\mathrm{cm^{-2}}$ (Obs2), showing no significant variation of the absorbing material between observations. 
For the electron temperature ($kT$), we obtain confidence ranges $0.71 - 0.85~\mathrm{keV}$ (Obs1) and $0.87 - 1.04~\mathrm{keV}$ (Obs2), implying a slightly warmer plasma for Obs2. Metal abundance ($Abund.$) is sub-solar and confidence ranges are $0.12 - 0.74$ for Obs1 and $0.07 - 0.16$ for Obs2.

Based on the results of the individual fit of each observation, and in order to increase the fit statistic, we simultaneously fit Obs1 and Obs2 spectra using a thermal plasma model with the same $N(\mathrm{H})$. 
In Table \ref{tabla_apec} we show the fitting results for the absorbed thermal plasma, considering two procedures: i) all parameters except normalisation are tie together between Obs1 and Obs2, and ii) only N$_H$ is tie together between the observations.

The results of the spectral fitting reported in Table \ref{tabla_apec} show that the X-ray emission from 4XMM J184650.6$-$022907 is likely produced by a thermal plasma with a temperature $\lesssim 1~\mathrm{keV}$ and sub-solar abundance. There is a slight increase in the plasma temperature between observations (refer to the right column of the Table), and no significant variations of the metal abundance. The normalisation increases by a factor $\sim 2$ between observations, i.e. in a time scale of 7 years. This indicates that the increase of the flux is produced by an increase of the amount of emitting plasma.  

\subsection{IRAS 18443$-$0231: a planetary nebula candidate}
\label{pnsect}

IRAS 18443$-$0231 was first catalogued as a planetary nebula (PN) by \citet{urquhart2009} as part of the RMS survey. \citet{anderson11}, in their Green Bank Telescope \hii~region survey proposed this source as a ``possible'' PN. 
Later, it was classified as a PN by \citet{cooper2013} from its near-infrared spectrum and the absence of millimetre dust continuum emission. Finally, in a near-infrared survey of the inner Galactic plane for Wolf-Rayet stars, \citet{kanarek2015} also suggested a PN nature for this source. 

The systemic velocity of IRAS 18443$-$0231 is 110 km s$^{-1}$ (see the RMS Database Server; \citealt{lumsden2013}). \citet{yang2016}, in their study of distances to PN based on HI absorption, determined the far distance as the most probable for our source, which correspond to a kinematic distance of about 8.2 kpc. By analysing the spectrum of the H$_2$O maser from \citet{urqu11}, we point out that this source could be a ``water-fountain'' PN, which is indeed a very interesting kind of object due to the scarce of them (e.g. \citealt{boboltz05,fan24}).  On the other hand, as it was mentioned in Section \ref{intro}, 4FGL-DR3 and 4FGL-DR4 Fermi catalogues refer to the radio source TXS 1844$-$025 (18:47:00.5, $-$02:27:52, J2000), which is indeed IRAS 18443$-$0231, as a `possible association'. Thus, we conclude that such a source deserves a thoroughest analysis in the context of this work. 

\begin{figure}
\centering
\includegraphics[width=7cm]{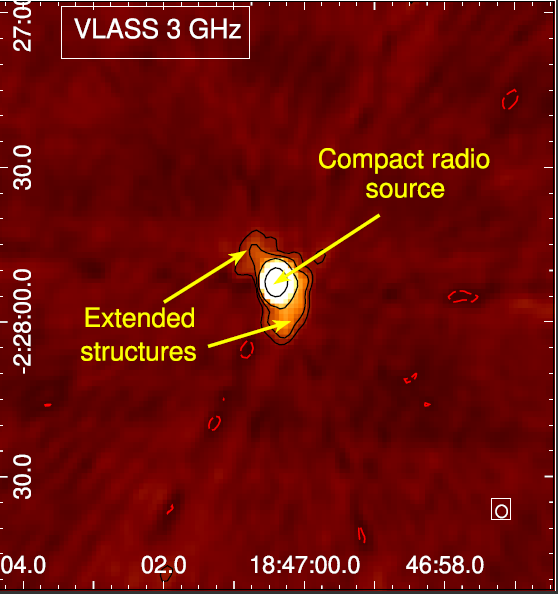}
\includegraphics[width=7cm]{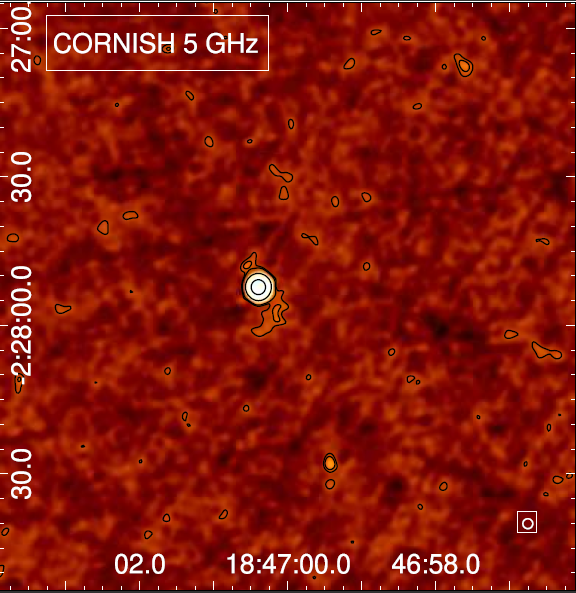}
\caption{Top panel: Colour-scale and black contours show the radio continuum emission at 3 GHz towards IRAS 18443$-$0231 extracted from VLASS. Levels are at 1 (5$\sigma$), 3, 8, and 50 mJy beam$^{-1}$. Red contour level is at $-$1 mJy beam$^{-1}$. Bottom panel:  Colour-scale and black contours show the radio continuum emission at 5 GHz towards IRAS 18443$-$0231 extracted from CORNISH. Levels are at 0.4 (5$\sigma$), 0.6, 5, and 50 mJy beam$^{-1}$. The beam are indicated at the bottom right corner.}
\label{PPN_VLASS}
\end{figure}

Figure\,\ref{PPN_VLASS}-top presents in colour and black contours the radio continuum emission at 3 GHz extracted from the second epoch of VLASS observations (VLASS 2.1) towards the PN candidate IRAS 18443$-$0231. It can be noticed a conspicuous compact radio source accompanied by a weak extended emission in the form of two not completely aligned lobes, which are clearly related to the central source. Figure\,\ref{PPN_VLASS}-bottom presents in colour and black contours the radio continuum emission at 5 GHz extracted from CORNISH. It can also be appreciated the conspicuous central source and interestingly, although weaker, the vestiges of the lobes detected more clearly at 3 GHz can also be seen. Given their morphology and that they appear in both images, we wonder if these structures could be jets emerging from the central object. The dominant radio continuum emission in PNe is expected to be thermal, although some processes such as jets and magnetic fields in very young PNe (protoplanetary nebulae) could provide an environment for non-thermal emission.

\begin{table}
\centering
\caption{Integrated radio flux densities at 1.4 and 5 GHz and the derived radio spectral index ($\alpha$; from S$_{\nu} \propto \nu^{\alpha}$) of IRAS 18443$-$0231.}
\begin{tabular}{cccc}
    \hline\hline
          S$_{\rm 1.4 GHz}$   & S$_{\rm 5 GHz}$ & Work &  Spectral index   \\
            (mJy)             &      (mJy)      &      &       ($\alpha$)      \\      
         \hline
           905.0$^{\bf *}$  & 552.4$^{\bf *}$ &  \citet{becker1994} & $-$0.390 $\pm$0.076$^{\dagger}$\\
           724.19 $\pm$4.18  & 349.87 $\pm$0.27 &  \citet{white2005}  & $-$0.570 $\pm$0.003 \\
          \hline\hline
\multicolumn{4}{l}{ * No reported errors.} \\
\multicolumn{4}{l}{$\dagger$ Assuming arbitrary and coarse errors of 5$\%$ in flux densities.}
\end{tabular}
\label{spectral_index}
\end{table}

\begin{figure}
\centering
\includegraphics[width=9cm]{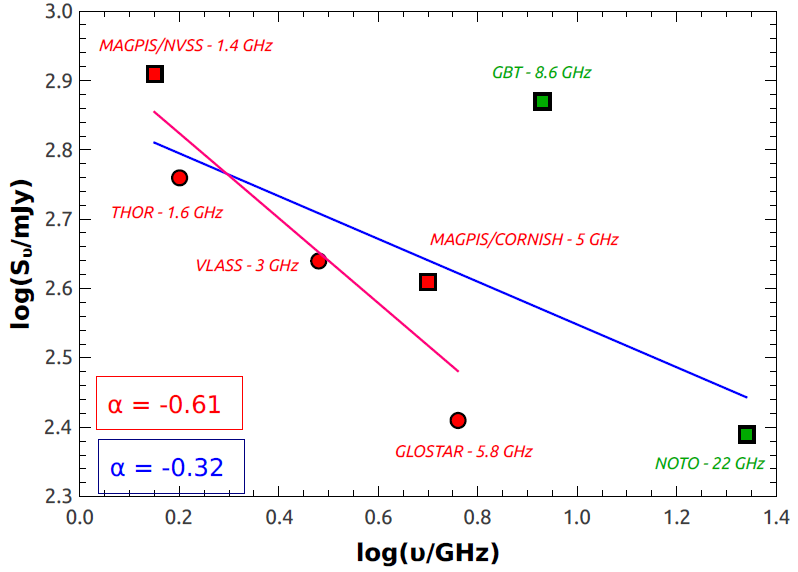}
\caption{Flux density as a function of frequency for IRAS 18443$-$0231.  Red squares indicate averaged flux densities extracted from several catalogues, red circles show a single flux density, and  green squares indicate flux densities obtained from single dish observations (see Table\,\ref{radio_fluxes}).  The flux density at 3 GHz was derived in this work. Error bars are contained in the point symbols. Rough spectral index estimations are presented:  blue and red lines are the best ﬁtting from a power law of the form S$_{\nu} \propto \nu^{\alpha}$ to all available data and to data without considering the single dish observations, respectively. It is important to mention that this is a rough and complementary analysis done with the aim of including all fluxes available in the literature. Possible variability, primary-beam correction, missing flux density effects are not taken into account. }
\label{PPN_RSI}
\end{figure}

Therefore, we investigated the nature of the radio continuum emission in terms of its spectral index. Based on the works/catalogues of \citet{zoone1990}, \citet{becker1994},  and \citet{white2005}, which give the values of the integrated flux densities at 1.4 and 5 GHz from observations carried out with the VLA radio interferometer, we determined the radio spectral index ($\alpha$; from S$_{\nu} \propto \nu^{\alpha}$) of the source. It is worth noting that the homogeneity of the observations of each work (angular resolution and interferometer configuration) makes completely reliable the obtained spectral index values. Table\,\ref{spectral_index} shows the integrated radio flux densities at 1.4 and 5 GHz (Cols.\,1 and 2, respectively) extracted from the Vizier catalogues associated with the works listed in Col.\,3. The estimated spectral index values (presented in Col.\,4.) are consistent with non-thermal radio continuum emission associated with the synchrotron mechanism. In the case of the data from \citet{becker1994} there are not reported errors for the flux densities, neither in the Vizier catalogue nor in the paper. Hence, we assumed an arbitrary and coarse error of 5$\%$ for such flux densities to estimate a likely upper limit for the error of the corresponding $\alpha$.

\begin{table*}
\centering
\caption{Summary of the available radio flux densities of IRAS 18443-0231.  }
\begin{tabular}{ccccc}
    \hline\hline
           Frequency & Telescope &    Beam    & S$_{\rm int}$ &  Work (Catalogue) \\
              (GHz)     &  &($\arcsec$) &     (mJy)     & \\
            
        \hline
       1.4 &  VLA (A and B conf.)     &  5  & 905 (NA) & \citet{zoone1990} \\
           &  VLA (D conf.)   & 45 & 800 $\pm$ 24 & \citet{condon1998} (NVSS)\\
           &  VLA (B and BnA conf.)   &  6  & 724 $\pm$ 4 & \citet{white2005} (MAGPIS)\\
           &  VLA (B, C, and D conf.)$^{*}$ &  6  & 678 $\pm$ 1 & \citet{helfand2006} (MAGPIS)\\
       \hline
       1.6 & VLA (C conf.) & 13$\sim$14 & 574.0 $\pm$ 1.3 & \citet{bihr2016} (THOR)\\
       \hline      
       3 &  VLA (B and BnA conf.) & 2.5 & 440 $\pm$ 23 & This work$^{**}$\\       
       \hline 
       5 & VLA (C and CnD conf.) &  4$\sim$5 & 552  (NA) &\citet{becker1994} (GPS5)\\
         & VLA (C and CnB conf.) & 6 & 349.87 $\pm$ 0.27 & \citet{white2005} (MAGPIS)\\
         & VLA (B and BnA conf.) & 1.5 & 360.35 $\pm$ 32.91 & \citet{purcell2008} (CORNISH)\\
       \hline  
       5.8  & VLA (B conf.)      &  1   & 261.52 $\pm$ 13.08 &  \citet{dzib2023}                 \\ 
       \hline
         8.6  & Green Bank Telescope    & 82    & 752 $\pm$ 41$^{***}$    & \citet{anderson11} \\ 
       \hline
        22  &  32 m INAF-IRA Noto Radiotelescope & 54 & 250 $\pm$ 60$^{***}$ & \citet{leto2009}\\
       \hline\hline
\multicolumn{5}{l}{* VLA images combined with the Effelsberg 100 m telescope ($\sim 9 \arcmin$ angular resolution).}\\
\multicolumn{5}{l}{** Using the second epoch of VLASS observations (VLASS 2.1) extracted from the Canadian Astronomy Data Centre.}\\
\multicolumn{5}{l}{*** Single dish observations whose flux densities could include extended emission contribution.}\\
\end{tabular}
\label{radio_fluxes}
\end{table*}

Even though the above presented results are pretty conclusive, additionally, we carried out a rough study of the spectral index based on all IRAS 18443$-$0231 radio continuum flux densities available in the literature (see Table\,\ref{radio_fluxes} and Fig.\,\ref{PPN_RSI}). With the caveat that in general such flux density points come from observations that were carried out with a variety of instruments from 1990 to 2023,  and  possible variability, primary-beam correction, and missing flux density effects are not taken into account, we present this complementary analysis as a support  of the above obtained results.  
To the available radio flux densities at different frequencies (Table\,\ref{radio_fluxes}) we added the flux density at 3 GHz estimated in this work. 
In the cases with several flux density values from different catalogues or works for the same frequency, the average of such values was considered (red squares in Fig.\,\ref{PPN_RSI}).  It is worth noting that it is possible that the points at 8.6 and 22 GHz (green squares in Fig.\,\ref{PPN_RSI}), obtained from single dish observations, could include extended emission not related to the source, and hence, such flux density values could be upper limits. Two linear fittings were done: one considering all points (blue line in Fig.\,\ref{PPN_RSI}) obtaining a  
 radio spectral index $\alpha$ ($S_{\nu} \propto \nu^{\alpha}$) of $-0.32$, and the other one, considering only the points from interferometric observations (red line in Fig.\,\ref{PPN_RSI}), 
 obtaining an $\alpha$ of $-0.61$. 

Finally, it is important to note that if we take into account only the flux densities obtained from the six sub-bands in the range 1--2 GHz of the THOR survey (observations acquired in July 2013), the spectral index yields a value of about $-0.1$ \citep{bihr2016}. In spite of that such spectral range is a portion of the range analysed here, this result could be indicating the preponderance of thermal emission during a certain period in IRAS 18443$-$0231. This could be explained by the existence of variability effects, indicating that the non-thermal emission could be a transitory phenomenon in this kind of objects (see Sect.\,\ref{ppndiscus} and \citealt{suarez15}).

Finding non-thermal radio continuum emission towards this type of objects is indeed an interesting and very uncommon result. The obtained negative spectral radio index very likely indicates synchrotron emission due to particle acceleration related to jets, strongly suggesting that IRAS 18443$-$0231, catalogued as a PN candidate, is in fact a protoplanetary nebula (pPN) (see Sect.\,\ref{ppndiscus}).

In order to better understand the nature of this object, we characterise the environment in which it is embedded or located. At submillimeter wavelengths, even though it is not catalogued as a source in ATLASGAL, Bolocam, or SCUBA, we observed a very faint structure at 850 $\mu$m emission from data of the last bolometer (see green contours in Fig.\,\ref{outflow}). This suggests that IRAS 18443$-$0231 is related to a weak ($S_{mean} \sim 0.08$ Jy beam$^{-1}$; 2$\sigma$ above rms noise) and small (slightly larger than the beam) molecular clump. 

\begin{figure}
\centering
\includegraphics[width=8.7cm]{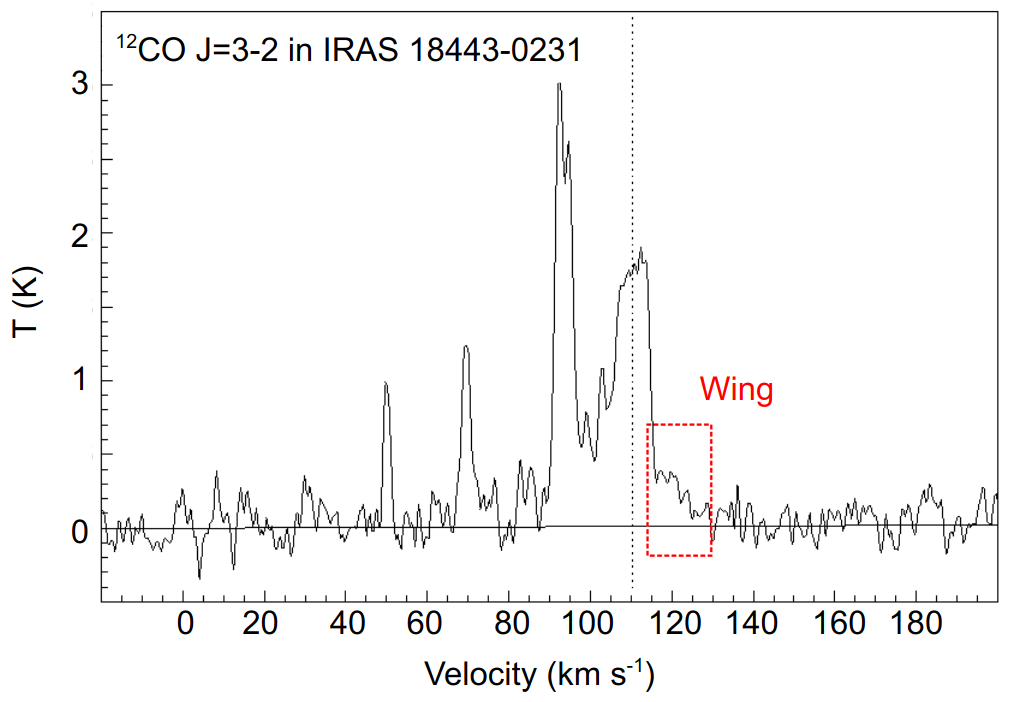}
\caption{$^{12}$CO J=3--2 spectrum obtained towards IRAS 18443-0231. The spectral wing indicating a red-shifted molecular outflow is marked. The vertical dotted line represents the systemic velocity of the source.}
\label{cospect}
\end{figure}

\begin{figure}
\centering
\includegraphics[width=7.5cm]{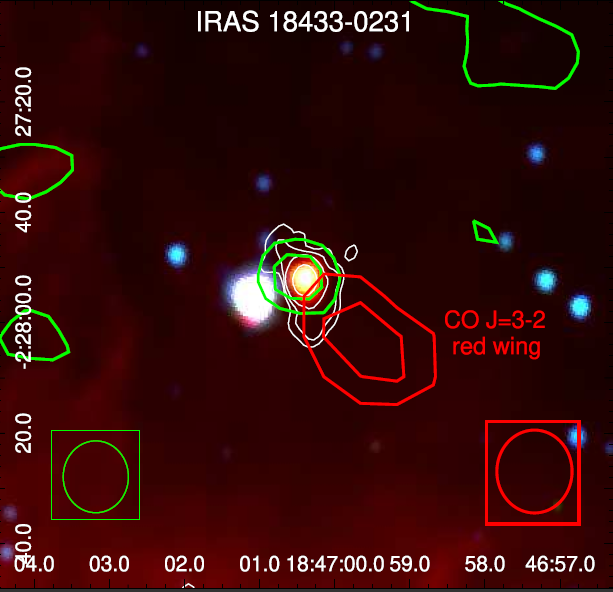}
\caption{Three-colour Spitzer image towards IRAS 18443$-$0231 with 3.6, 4.5, and 8 $\mu$m emissions represented in blue, green, and red, respectively. The white contours are the same ones that were presented in Fig.\,\ref{PPN_VLASS}. Green contours represent the continuum emission at 850 $\mu$m extracted from SCUBA. Levels are at 0.072 and 0.087 Jy beam$^{-1}$. The red contours represent the $^{12}$CO J=3--2 red wing integrated between 115 and 130 km\,s$^{-1}$ (see Fig.\,\ref{cospect}) with levels are at 9 and 12~K\,km\,s$^{-1}$.}
\label{outflow}
\end{figure}

Additionally, we looked for molecular outflows towards this source. Using data from the JCMT we found a spectral wing in the $^{12}$CO J=3--2 spectrum suggesting the presence of a red-shifted outflow (see Fig.\,\ref{cospect}). It can be noticed that if there was a blue-shifted outflow it would be masked within the $^{12}$CO component centred at 95 km\,s$^{-1}$. Figure\,\ref{outflow} displays in red contours the $^{12}$CO emission integrated between 115 and 130 km\,s$^{-1}$ (i.e. along the spectral wing shown in Fig.\,\ref{cospect}) and clearly shows the presence of a lobe corresponding to a red-shifted molecular outflow emanating from IRAS 18443-0231.

\subsubsection{X-ray analysis of IRAS 18443$-$0231}

\begin{figure}
\centering
\includegraphics[width=8cm]{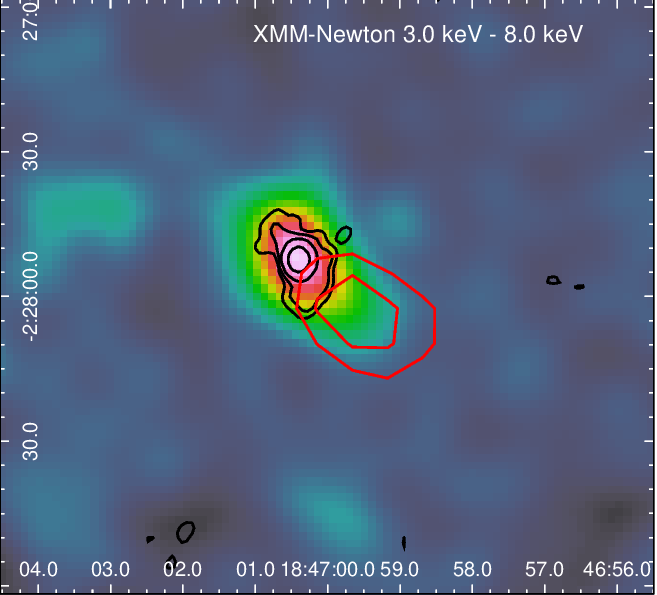}
\caption{X-ray emission between 3.0 and 8.0 keV associated with the source 4XMM J184700.4$-$022752 (IRAS 18443$-$0231) displayed in colour. The black contours correspond to the 3 GHz radio continuum emission related to IRAS 18443$-$0231. Levels are at 1 (5$\sigma$), 3, 8, and 50 mJy beam$^{-1}$. The red contours represent the $^{12}$CO J=3--2 red wing integrated between 115 and 130 km\,s$^{-1}$ (see Fig.\,\ref{cospect}) with levels are at 9 and 12 K\,k\,ms$^{-1}$. }
\label{3GHz-Xray}
\end{figure}

Figure\,\ref{3GHz-Xray} shows the good spatial correlation between the radio continuum emission at 3 GHz associated with IRAS 18443$-$0231 and the X-ray source 4XMM J184700.4$-$022752. Moreover, it can be appreciated a diffuse extended structure towards the southwest in the X-ray emission, which seems to accompany the direction of the red lobe of the $^{12}$CO molecular outflow. 

4XMM J184700.4$-$022752 is a hard X-ray source, with most of the photons detected between 2.0 and $6.0~\mathrm{keV}$. The reported count rate in the EPIC-pn cameras are $1.65\pm0.33 \times 10^{3}~\mathrm{cts\,s^{-1}}$ and $1.49\pm0.24 \times 10^{3}~\mathrm{cts\,s^{-1}}$ for Obs1 and Obs2, respectively, indicating no significant variability between observations (refer to the 4XMM-DR13 catalogue).
To analyse the X-ray emission from 4XMM J184700.4$-$022752, we extracted source and background spectra from the 6 cameras. We obtained 289 source counts, which in principle is sufficient for a spectral fitting. Taking into account the lack of variability, we combined the source spectra of the 6 cameras into a single spectrum using the {\it epicspeccombine} task of SAS. The merged spectrum was binned with a minimum of 15 counts per bin. We used Xspec to display and analyse the spectrum. We fitted the spectrum using different emission models to determine whether the emission has a thermal or non-thermal nature.  

\begin{figure}
\centering
\includegraphics[width=6.3cm,angle=-90]{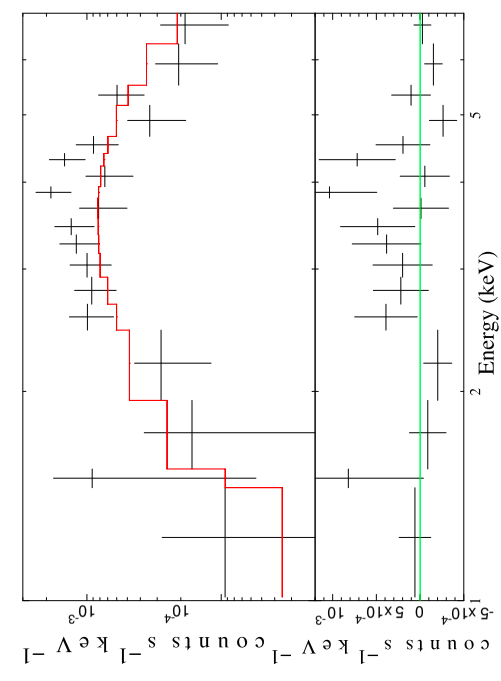}
\caption{X-ray spectrum of 4XMM J184700.4$-$022752 between 1.0 and $7.0~\mathrm{keV}$, obtained by combining the 6 individual spectra of Obs1 and Obs2. The best-fit absorbed power law for a photon index $\Gamma_X = 1.5$ is shown in red. }
\label{PNN_X}
\end{figure}

Unfortunately, all the different models we tried were not well constraint and the parameters had large uncertainties.
Then, for the purpose of estimating the X-ray flux of the source, we assumed some spectral behaviour. 
Taking into account that the radio emission of the pPN is likely produced by synchrotron radiation and has a spectral index $\alpha \sim -0.5$, we considered an absorbed power law ($tbabs \times powerlaw$) in the X-ray band with a fixed photon index $\Gamma_X = 1 - \alpha = 1.5$. 
In other words, we assumed that the non-thermal radio emission extends to the X-ray domain without a spectral break. We fitted the spectrum on the $1.0-7.0~\mathrm{keV}$ energy band letting the column density of the absorbing material $N(\mathrm{H})$ and normalisation $Norm$ as free parameters. We obtained $N(\mathrm{H}) = 5.4^{+1.7}_{-1.3} \times 10^{22}~\mathrm{cm^{-2}}$, $Norm=2.45^{+0.58}_{-0.32}$ and $\chi_r^2=1.02$ for 19 d.o.f. In Fig.\,\ref{PNN_X} we show the merged spectrum and the best fit power law model. The unabsorbed flux in the $1.0-7.0~\mathrm{keV}$ band is $1.3\pm0.3 \times 10^{-13}~\mathrm{erg\,s^{-1}\,cm^{-2}}$ and the luminosity for a distance of $8~\mathrm{kpc}$ is $9.9\pm2.0 \times 10^{32}~\mathrm{erg\,s^{-1}}$. This value is compatible with the expected non-thermal X-ray luminosity in very young pPNe \citep{blackman2001,sahai2003}.  

\section{Discussion about the nature of the $\gamma$-ray emission}
\label{discussion}

In this section, the nature of the $\gamma$-ray source 4FGL J1846.9$-$0227 is discussed based on the previous multiwavelength characterisation of the region.

\subsection{Is 4FGL J1846.9$-$0227 a blazar?}

\citet{kerby21} analysed a sample of 174 unidentified Fermi sources with a single X-ray/UV/optical counterpart. Given that the majority of the sources of 4FGL catalogue are pulsars and blazars, the authors developed a neural network classifier approach using $\gamma$-ray, X-ray, and UV/optical spectral parameters to yield a descriptive classification of unassociated spectra into pulsars and blazars. The study included the Fermi-LAT source 4FGL J1846.9$-$0227, but for the reasons mentioned in Sect.\,\ref{intro}, the association of this $\gamma$-ray source with a blazar was ambiguous,  which encouraged us to carry out a more in-depth study  of the source.

First of all, the location of 4FGL J1846.9$-$0227, close to the galactic plane ($b \sim -0\fdg16$), would suggest a galactic origin for the high energy source.  
 Then, as shown above, we discarded the blazar nature of the the X-ray sources indicated by \citet{kerby21}. It is worth noting that the X-ray emission of the most intense one, that it is related to 4XMM J184650.6$-$022907, was modelled by an optically thin thermal plasma (see Sect.\,\ref{spec_X}). This indeed discards a blazar as a possible counterpart, given that these extra-galactic sources typically produce non-thermal X-rays via the synchrotron mechanism \citep{kaur2019,kerby21}.  

 In conclusion, 4XMM J184650.6$-$022907 related to a binary system and the source IRAS 18443-0231 are the best candidates to focus our attention to try to explain the high-energy emission.

\subsection{Relating 4XMM J184650.6$-$022907 with the binary star BD$-$02 4739}

Having ruled out a blazar as  possible counterpart of 4XMM J184650.6$-$022907, the connection between the X-ray source and the spectroscopic binary star BD$-$02 4739 is important in the context of this work. As mentioned before, Fig.\,\ref{x+bd} shows the  very good spatial correlation between BD$-$02 4739 and 4XMM J184650.6$-$022907, which strongly suggests the connection between both sources. In what follows, we explore different scenarios to explain the X-ray emission arising from the binary star.  

\subsubsection{X-ray binaries}

X-ray binaries (XRBs) are binaries system composed by an accreting compact object, a neutron star or a black hole, orbiting a companion, or donor star, from which it accretes material. This material is accelerated in the strong gravitational field of the compact object and heated up to about $10^7$ K before being accreted, giving as a result the observed X-ray radiation. The basic division of XRBs into the high-mass (HMXBs) and low-mass (LMXBs) systems depends on the mass of the donor star. While the donor of LMXBs is a low-mass star, typically with masses $\lesssim$ 1 \msol, the masses of donor stars in HMXBs systems are $\gtrsim$ 10 \msol. 

In most HMXBs, the compact object usually is a neutron star, which accretes material from the stellar wind of its massive companion \citep{liu2006}. High-mass stars have powerful stellar winds with mass loss rates $\gtrsim$ 10$^{-6}$ \msol yr$^{-1}$ \citep{smith2014}, so direct wind accretion usually results in high accretion rates, producing X-ray luminosities in the 10$^{35} - 10^{40}$ erg s$^{-1}$ range \citep{forna2023}. The donor star in most HMXBs is either a supergiant O/B star or a Be star, both of which have higher mass loss rates than O/B main sequence stars. The HMXBs typically have bright X-ray hard spectrum, which can be adequately modelled by a power law \citep{haberl2022,sidoli2018}, however, some of them exhibit a small excess of soft X-rays whose origin is not entirely clear \citep{vander2005}. Therefore, taking into account that the X-ray luminosity of 4XMM J184650.6$-$02290 is about 10$^{30}$ erg s$^{-1}$, that its soft X-ray spectrum is well fitted by an optically thin thermal plasma model, and finally, that the UV/optical/IR SED indicates that the donor companion is a low-mass giant star, we can rule out that the binary BD$-$02 4739 is a HMXB.   
 
On the other hand, LMXBs are typically dimmer than HMXBs, with X-ray luminosities that can be as low as $10^{30}$ erg s$^{-1}$ for quiescent systems in which little or no accretion occurs \citep{bah2023}. Therefore, at least from the X-ray luminosity point of view, we cannot rule out that 4XMM J184650.6-022907 is a LMXB.  Moreover, the characteristics of the donor star (a low-mass giant star) are consistent with LMXBs. However, 
X-ray emission from LMXBs is not expected to be produced by an optically thin thermal plasma, as we found for 4XMM J184650.6$-$022907. The spectral emission of LMXBs can be composed of three main components: disk blackbody originating from optically-thick accretion disc, a comptonized component originating due to comptonization of X-rays in the corona, and eventually, a blackbody component of emission from the neutron star surface or the boundary layer \citep{lin2009}. Regarding energy spectra modelling of LMXBs, there are two different approaches: (1) a soft/thermal, and (2) a hard/comptonized components \citep{mitsuda1989, hasinger1989}, typically fitted by blackbody and power-law models \citep{abdel2021}, respectively. In summary, given the behaviour of the X-ray spectrum of 4XMM J184650.6$-$022907 it is unlikely that the X-ray emission is originated in a LMXB.

\subsubsection{A symbiotic binary system}
\label{symbiotic}

We investigated the possibility that the X-ray emission is being originated by a symbiotic binary system. Such kind of sources are wide binary systems in which a compact object, usually a white dwarf (WD), accretes from a more evolved companion, typically a red giant. 

The bolometric luminosity of an accreting WD with a red giant companion is set mainly by whether or not accreted material burns quasi-steadily in a shell on the surface of the WD. 
During the shell burning stage the luminosity of the WD is typically 10$^3$ L$_{\odot}$, but if shell burning is absent, its luminosity is of the order of 10 L$_{\odot}$ \citep{soko2016}. Taking into account that in absence of shell burning, the red giant, and in some cases the accretion disk, dominates the optical spectrum \citep{soko2016}, the bolometric luminosity derived in Sect.\,\ref{SED} for BD$-$02 4739 star of about 23 L$_{\odot}$ is consistent with a scenario of a WD-symbiotic with non shell burning. 

It is worth noting that the X-ray emission from 4XMM J184650.6$-$022907 is also compatible with the so-called $\beta$-type symbiotics, first introduced by \cite{muerset97}. This class is characterised by soft emission produced by photons with energies less than $\sim 2-3\,\mathrm{keV}$. According to \cite{soko2016}, both burning and non-burning symbiotics may belong to the $\beta$-type, in which thermal X-ray emission from an optically thin plasma arises from the gas heated by colliding winds and/or collimated jets.

Typical X-ray luminosities of $\beta$-type symbiotics in the $0.3-2.4~\mathrm{keV}$ band that goes from $10^{30}$ to $10^{31}~\mathrm{erg\,s^{-1}}$ \citep{luna2013}. The luminosities of 4XMM J184650.6$-$022907 in this energy band are $\sim 4.8\times 10^{30}$ and $\sim 9.0\times 10^{30}~\mathrm{erg\,s^{-1}}$ for Obs1 and Obs2, respectively, estimated with the best-fit model of Table\,\ref{tabla_apec} (right column).   
A striking characteristic of the X-ray emission from 4XMM J184650.6$-$022907 is its variability, namely the flux increases by factor $\sim 2$ between 2011 and 2018. A similar behaviour has been observed in AG Peg, a $\beta$-type symbiotic stars that presented an outburst in 2015. This star has been extensively studied in the X-ray domain since its discover by ROSAT in 1993. It was observed with Swift in 2013 and 2015-2016, and with XMM-Newton in 2017. The X-ray emission, which was modelled by an optically thin plasma, increased by a factor 2 during the active phase in 2015, with respect to the pre- and post-outburst fluxes \citep{zhekov18}. 
While the X-ray emission in the pre-outburst period is likely produced by plasma heated in the colliding-wind region of the binary system, the dominant heating mechanism during the active phase is probably accretion of the wind from the cool star into the WD \citep{zhekov16} or, alternatively, shocks from the interaction between material ejected during the outburst and the ejecta and/or circumbinary material \citep{ramsay16}.

Regarding the radio continuum emission towards symbiotic stars, seminal studies like \citet{sea1993} found  that most shell burning symbiotics detected with the Very Large Array (VLA) had radio flux densities on the order of mJy, consistent with free-free emission from the ionised wind of the red giant. On the other hand, \citet{weston2016}, studying a sample of 11 non-burning symbiotics with radio continuum observations, found that about half of the sources have faint radio flux densities of about 10 $\mu$m or less. Taking this into account, we searched for possible radio counterpart for the binary star BD$-$02 4739 using the 3 GHz radio continuum  Very Large Array Sky Survey \citep[VLASS;][]{lacy2020}. Although the high sensitivity ($\sim 100~\mu$Jy beam$^{-1}$) and the high resolution ($\sim 2\arcsec$) of the data, we did not find any evidence of radio continuum emission related to the source, which is consistent with a scenario of non-burning WD-symbiotic star.

Summarising, our multiwavelength analysis strongly suggests that 4XMM J184650.6$-$022907 is related to the spectroscopic binary star BD$-$02 4739, which would be a non-burning $\beta$-type WD-symbiotic system with a giant donor star. 
The increase of the X-ray flux detected in the 2018 XMM-Newton observation could be a consequence of  an active phase transited by the system.  As pointed out by \cite{soko2016}, in a non shell burning symbiotic, the optical emission is dominated by the red giant. This supports our hypothesis of Sec.\,\ref{SED}, where we consider that the fluxes used to fit the SED come from the donor star.

\subsection{Is BD$-$02 4739 responsible for the $\gamma$-ray emission of 4FGL J1846.9$-$0227?}

Symbiotic stars have been confirmed as potential $\gamma$-ray sources in the GeV band, when they experience a nova eruption event \citep{li2020, fran2018, abdo2010}. Novae are thermonuclear explosions on a WD surface fuelled by mass accreted from a companion star. Symbiotic novae have an evolved companion (e.g. red giant) with a dense wind as opposed to a main-sequence companion for classical novae \citep{ackermann14}. Current physical models suggest that shocked expanding gas from the symbiotic nova shell interacts with the dense wind of the red giant, and that particles can be accelerated effectively to produce $\pi^{0}$ decay $\gamma$-rays from proton-proton interactions \citep{fran2018}.
The first Fermi-LAT $\gamma$-ray detection of a nova was made in the symbiotic-like nova V407 Cyg in 2010 \citep{abdo2010}, followed by few other examples such as V1535 Sco, V1534 Sco, and V745 Sco.

\begin{figure}
\centering
\includegraphics[width=8.7cm]{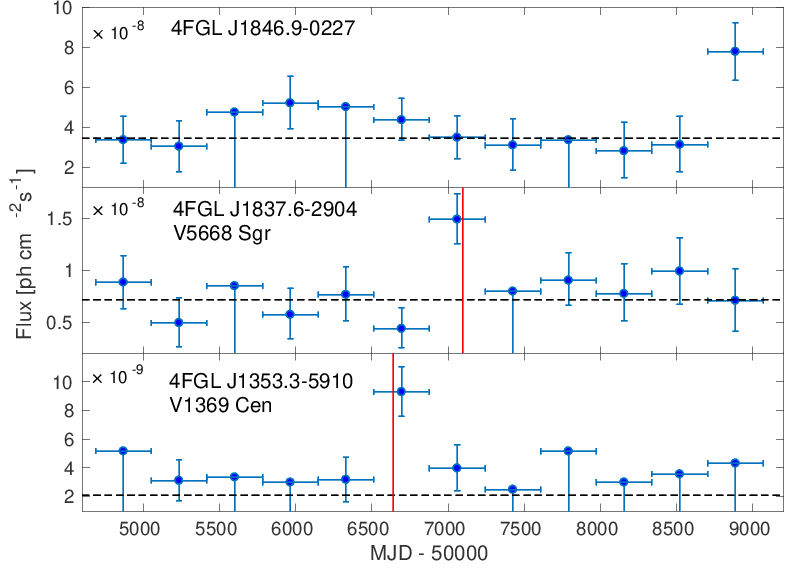}
\caption{12 years light curves of Fermi, for 4FGL J1846.9$-$0227 (top), V5668 Sgr (middle), and V1369 Cen (bottom). Each temporal bin corresponds to 1 year of data (starting August 2008). The horizontal dotted line is the time-averaged flux and the vertical red lines indicate the maximum of the optical novae. Fluxes with significance $\sigma~(=\sqrt{TS}) < 2$ are shown as 95\% confidence upper limits, where $TS$ is the test statistics reported in the 4FGL-DR3 catalogue for each epoch.}
\label{fig_gamma}
\end{figure}

We constructed a light curve of the GeV emission of 4FGL J1846.9$-$0227 using data from the Incremental Fermi LAT 4th Source Catalogue (4FGL-DR3), spanning 12 years of data (August 2008 - August 2020) with 1-year bins (Fig.\,\ref{fig_gamma}). For comparison, we also shown the light curves of novae V5668 Sgr (4FGL J1837.6$-$2904) and V1369 Cen (4FGL J1353.3$-$5910), where we have marked the date of the optical maximum. \cite{fran2018} show that gamma-ray peak of novae is concurrent with the optical maximum and lasts for some tens of days. Thus, we wonder if BD$-$02 4739 experienced a nova event.

Although the source experiences an increase in the $\gamma$-rays photon flux between August 2019 and August 2020, it is important to note that it presents a considerably continuous $\gamma$-ray emission along the years, which is not compatible with the fact that classical and symbiotic novae are only found to be GeV-emitters during the
outbursts. In addition, we have not found any near-infrared/optical correspondence (VISTA, ASAS, ASAS-SN, Gaia databases) that shows a nova event. It is worth noting that based on the optical magnitude of BD-02 4739, it would be very difficult that the putative nova has been missed by both professional and amateur observatories. Moreover, by considering its nearby distance (0.5 kpc), a nova event, if not missed by Fermi, it would be extremely brighter in $\gamma$-rays than what it is shown in Fig.\,\ref{fig_gamma} (upper panel); for instance see the RS Oph 2021 case at a distance of about 2.5 kpc \citep{cheung22}.

In conclusion, in spite of these interesting and new results about this binary system, we discard it as the responsible of the detected GeV emission in the region. 

\subsection{What about MYSO MSX G30? }

From our analysis based on ALMA data, we found that the confirmed MYSO MSX G30 is associated with a dense molecular core, which presents evidence of star formation activity due to the presence of numerous molecular lines (Fig.\,\ref{spw3}), which are characteristic of hot molecular cores. This chemical richness is an interesting issue to be studied in deep in a future work. In particular, what is most relevant to this study is the detection of formaldehyde molecular outflows (see Fig.\,\ref{zero-moments}), which could explain the $\gamma$-ray emission of the Fermi-LAT source, as it was proposed by \citet{munar2011}. However, based on the position of MYSO MSX G30 at the border of the confidence ellipse of the Fermi-LAT source, and the fact that the MYSO does not present any X-ray or centimeter radio emissions associated, we suggest that, it is very likely that the molecular outflows detected towards this MYSO are not energetic enough to be responsible for the $\gamma$-ray emission. Therefore, we conclude that it is very unlikely that the counterpart of the high-energy source is such a young protostar.

\subsection{Is IRAS 18443$-$0231 generating the $\gamma$-ray emission of 4FGL J1846.9$-$0227? }
\label{ppndiscus}

The source morphology at 3 and 5 GHz that we shown in Fig.\,\ref{PPN_VLASS} together
with the non-thermal nature of the radio continuum emission, which is in line with 
what was indicated by \citet{irabor18}, strongly suggest the presence of jets and accelerated particles in IRAS 18443$-$0231. Additionally, we discovered a red-shifted molecular outflow supporting the presence of such jets arising from the central source, and moreover, the X-ray morphology seems to encompass the outflow extension. The presence of jets and molecular outflows are in agreement with the detected H$_{2}$O maser by \citet{urqu11}. 

Taking into account that this source was previously catalogued as a PN candidate, our results indicate that IRAS 18443$-$0231 actually is a protoplanetary nebula (pPN), which is a short-lived transition (about $10^3$ yr) from the asymptotic giant branch (AGB) to the PN phase. There are not much cases that non-thermal radio continuum emission was observed towards this kind of objects (see the case of IRAS 15103-5754; \citealt{suarez15}). On the other side, it is more common observing molecular outflows in pPNe \citep{cox00,lore21}.
The non-thermal radio emission in pPNe suggests that the emitting electrons arise at collisions between the fast and slow AGB winds that are observed predominantly on the front sides of the circumstellar shells \citep{bains09}.   

Having discarded all the sources presented above as responsible of the $\gamma$-ray emission, taking into account the location of IRAS 18443$-$0231 (almost at the centre of the Fermi confidence ellipse), and mainly considering the presence of non-thermal jets accelerating particles, we wonder if IRAS 18443$-$0231 could be the counterpart of 4FGL J1846.9$-$0227. 

As mention by \citet{suarez15} for IRAS 15103$-$5754, the radio spectral index $\alpha$ between $-0.39$ and $-0.57$ measured in IRAS 18443-0231 is similar to those
found in synchrotron emission in galaxies with active nuclei. Following the authors, in the energy conditions of PNe, mass-loss processes can produce collisionless, non-relativistic shocks, and electrons may undergo diffusive shock acceleration up to very high velocities (`first-order Fermi acceleration').  In such a scenario, in which particles are accelerated, and in addition the source seems to be related to some surrounding molecular molecular gas (see Sect.\,\ref{pnsect}), processes such as proton-proton collisions and relativistic Bremsstrahlung are likely to occur generating $\gamma$-rays.  

Additionally, \citet{suarez15} noted that the PN IRAS 15103-5754 presents non-thermal radio emission and a spectral index that has become flatter in the lapse of a few years. They attribute this flattening to the arising of a photoionized nebula.
In our case, the radio spectral index obtained only from the THOR dataset towards IRAS 18443$-$0231 also could be revealing an arising of thermal emission. 
This would indicate that the very-high energy emission from pPNe should be a transient phenomenon.

In conclusion, we strongly suggest that IRAS 18443$-$0231 would be the responsible of the $\gamma$-ray emission. 
If this is the case, it would be the first reported pPN related to very high energy emission, and hence, multiwavelength dedicated observations, a follow up radio observations towards IRAS 18443$-$0231 and modelling are necessary to understand the mechanisms of $\gamma$-rays production in such a kind of source. 

\section{Summary and concluding remark} 

In previous works the nature of the Fermi-LAT source 4FGL J1846.9$-$0227 was discussed. In one of them it was suggested to be a blazar, and in other one it was mentioned the presence of a massive protostar within the Fermi confidence ellipse suggesting a possible association. Considering this discrepancy, in this work, we analysed several counterpart candidates to the $\gamma$-ray source based on a multiwavelength analysis.

First of all, from JVLA data we determined the thermal nature of an uncatalogued extended radio source in the field, which indicates that it should be an \hii~region and not a supernova remnant, ruling out that such a source is responsible for the $\gamma$-ray emission.
Using ALMA data, we confirmed that MSX6C G030.1981$-$00.1691 is a massive protostar with molecular outflow activity. However, since its peripheral location respect to the Fermi 95\% confidence ellipse, and the lack of associated X-ray and centimeter radio continuum emissions, it is very unlikely that this massive protostar is responsible of the $\gamma$-ray emission. 

We found that sources SwXF4 J184651.6-022507 and SwXF4 J184650.7-022904, previously suggested to be blazars, do not meet the mid-IR colour-colour criteria for this kind of sources. Additionally, they lie almost at the galactic plane, which weakens the possibility of a blazar detection. Moreover, the second one is related to the 4XMM 184650.6$-$022907 source, and based on XMM-Newton data, we found that this X-ray source, which lies closer to the centre of the Fermi 95\% confidence ellipse, is clearly associated with the spectroscopic binary star BD-02 4739. The X-ray spectrum of 4XMM 184650.6$-$022907 was well fitted by an optically thin thermal plasma model with a relatively low X-ray luminosity of about 10$^{30}$ erg s$^{-1}$. The soft thermal nature of the X-ray spectrum discards that 4XMM 184650.6$-$022907 is associated with a blazar. In fact, the X-ray spectrum is consistent with a white-dwarf (WD) symbiotic system. 
Thus, taking into account the results of the analysis of the X-ray data and the optical/infrared SED, we conclude that BD$-$02 4739 is a non-burning shell $\beta$-type WD symbiotic system with a giant donor star. The only way that this source could be considered as the responsible for the $\gamma$-ray emission is that it had experimented a nova event. 
Despite of these new and interesting results about this source, the absence of an optical light curve indicating a nova event strongly suggests that BD-02 4739/4XMM 184650.6$-$022907 is not associated with the high energy emission.

Finally, we found that IRAS 18443-0231, catalogued as a PN candidate located at a distance of about 8 kpc, would be actually a protoplanetary nebula. IRAS 18443-0231 is the closest source in the region to the centre of the Fermi 95\% confidence ellipse. We discovered extended and faint features related to the central peak of IRAS 18443$-$0231 at 3 and 5 GHz continuum emission and confirmed the non-thermal nature of the radio continuum (spectral index $\alpha$ between $-0.39$ and $-0.57$). These findings indicate  synchrotron emission very likely due to particles accelerated in jets. Additionally, we discovered a molecular outflow, supporting the presence of such jets, and we found that the source is surrounded by some molecular gas (traced by faint dust emission). IRAS 18443$-$0231 also presents hard X-ray extended emission, with the same inclination as the molecular outflow. 

Considering all the presented results, we conclude that the most promising source to be associated with 4FGL J1846.9$-$0227 is IRAS 18443$-$0231. The detection of synchrotron emission indicates that particles are accelerated at high velocities in the region, and considering the presence of some surrounding molecular gas, we suggest that processes such as proton-proton collisions and relativistic Bremsstrahlung are likely to occur.
If this is the case, it would be the first reported protoplanetary nebula related to very high energy emission, and hence, multiwavelength dedicated observations and modelling are necessary to understand the mechanisms of $\gamma$-rays production in this kind of source.

\section*{Acknowledgments}

We thank the referee Dr. J. Mart\'\i~for his comments and suggestions that helped us to improve this work.
M.O., A.P., and S.P. are members of the Carrera del Investigador Cient\'\i fico of CONICET, Argentina.  
This work was partially supported by the Argentina grants PIP 2021 11220200100012 and PICT 2021-GRF-TII-00061 awarded by CONICET and ANPCYT. 
This work is based on the following ALMA data: ADS/JAO.ALMA $\#$ 2015.1.01312. ALMA is a partnership of ESO (representing its member states), NSF (USA) and NINS (Japan), together with NRC (Canada), MOST and ASIAA (Taiwan), and KASI (Republic of Korea), in cooperation with the Republic of Chile. The Joint ALMA Observatory is operated by ESO, AUI/NRAO and NAOJ.
This publication makes use of VOSA, developed under the Spanish Virtual Observatory (https://svo.cab.inta-csic.es) project funded by MCIN/AEI/10.13039/501100011033/ through grant PID2020-112949GB-I00.
VOSA has been partially updated by using funding from the European Union's Horizon 2020 Research and Innovation Programme, under Grant Agreement nº 776403 (EXOPLANETS-A).

\section*{DATA AVAILABILITY}

The data underlying this paper were accessed from:
\begin{itemize} 
\small
\item \url{https://third.ucllnl.org/gps/}
\item \url{https://cornish.leeds.ac.uk/public/img_server.php} 
\item \url{https://irsa.ipac.caltech.edu/Missions/spitzer.html} 
\item \url{https://www.cadc-ccda.hia-iha.nrc-cnrc.gc.ca/en/search/}
\item \url{https://almascience.nrao.edu/aq/} 
\item \url{https://glostar.mpifr-bonn.mpg.de/glostar/image_server} 
\item \url{https://fermi.gsfc.nasa.gov/science/eteu/catalogs/} 
\item \url{https://thorserver.mpia.de/thor/image-server/}
\item \url{https://https://www.cosmos.esa.int/web/xmm-newton/xsa}
\end{itemize}

\noindent The catalogues underlying this paper were accessed from:
\begin{itemize}
\small
\item \url{https://vizier.cds.unistra.fr/viz-bin/VizieR}
\end{itemize}



\bibliographystyle{mnras}
\bibliography{ref} 

\end{document}